\documentclass[]{interact}

\usepackage{epstopdf}
\usepackage[caption=false]{subfig}

\usepackage{soul} 
\usepackage{xcolor} 
\usepackage[numbers,sort&compress]{natbib}
\bibpunct[, ]{[}{]}{,}{n}{,}{,}
\renewcommand\bibfont{\fontsize{10}{12}\selectfont}
\usepackage{multirow}

\theoremstyle{plain}

\theoremstyle{definition}

\usepackage{amsmath}
\usepackage{algorithm}
\usepackage{graphicx}
\usepackage{float}
\usepackage{hyperref}
\usepackage{algorithm}
\usepackage{algorithmic}
\usepackage{caption}
\usepackage{mathtools}
\usepackage{amsfonts}
\usepackage{comment}

\usepackage{array}
\usepackage{makecell}
\usepackage{multirow}

\theoremstyle{remark}

\begin{document}

\title{Masked Sequence Autoencoding for Enhanced Defect Visualization in Active Infrared Thermography}

\author{
\name{Mohammed Salah \textsuperscript{a} \thanks{CONTACT Yusra Abdulrahman. Email: yusra.abdulrahman@ku.ac.ae}, Eman Ouda \textsuperscript{a}, Stefano Sfarra \textsuperscript{b}, Davor Svetinovic \textsuperscript{c, d}, Yusra Abdulrahman \textsuperscript{a, e}}
\affil{\textsuperscript{a}Khalifa University of Science and Technology, Department of Aerospace Engineering, Abu Dhabi, United Arab Emirates \\ 
\textsuperscript{b}Department of Industrial and Information Engineering and Economics (DIIIE), University of L’Aquila, L'Aquila I-67100, Italy \\
\textsuperscript{c}Department of Computer Science, Khalifa University of Science and Technology, Abu Dhabi, UAE \\
\textsuperscript{d}ADIA Lab, Abu Dhabi, UAE \\ 
\textsuperscript{e}Advanced Research \& Innovation Center, Khalifa University of Science and Technology, Abu Dhabi, UAE}
}

\maketitle

\begin{abstract}
Active infrared thermography (AIRT) became a crucial tool in aerospace non-destructive testing (NDT), enabling the detection of hidden defects and anomalies in materials by capturing thermal responses over time. In AIRT, autoencoders are widely used to enhance defect detection by reducing the dimensionality of thermal data and improving the signal-to-noise ratio. However, traditional AIRT autoencoders often struggle to disentangle subtle defect features from dominant background responses, leading to suboptimal defect analysis under varying material and inspection conditions. To overcome this challenge, this work proposes a Masked CNN-Attention Autoencoder (AIRT-Masked-CAAE) that integrates convolutional feature extraction with attention mechanisms to capture both local thermal patterns and global contextual dependencies. The AIRT-Masked-CAAE introduces a masked sequence autoencoding strategy, where the network learns to infer missing thermal responses from surrounding contextual cues, while suppressing background redundancy. In addition, the proposed masked sequence autoencoding approach enables training on only a subset of the thermal sequence, while providing generalizable latent representations and reducing training time by a factor of 30. The AIRT-Masked-CAAE framework was evaluated using specimens made of PVC, CFRP, and PLA. The results demonstrate that the AIRT-Masked-CAAE surpasses state-of-the-art AIRT autoencoders in terms of contrast, signal-to-noise ratio (SNR), and metrics based on neural networks.
\end{abstract}


\section{Introduction}

As demand continues to grow for strict quality control, the aerospace industry has increased the adoption of advanced non-destructive testing (NDT) techniques during both the manufacturing and maintenance stages \cite{csk_nvs, ndt_review}. NDT enables the evaluation of material properties, structural integrity, and subsurface defects without causing damage to the component being tested. This makes it crucial in safety-critical domains such as aerospace. A broad range of NDT technologies, including ultrasonic testing, radiographic inspection, and infrared thermography (IRT), have been developed to detect hidden anomalies and subsurface defects in aerospace materials \cite{Tai2025}. Among these technologies, IRT has attracted considerable attention due to its fast scanning speed, non-contact nature, and cost-effectiveness \cite{irt_survey}. Recent studies have highlighted both the versatility of IRT in various materials \cite{helvig2025automated,zhang2025automatic,oswald2025inductive} and the growing dependence on advanced thermographic data processing techniques to overcome noise and improve defect visibility \cite{hsiao2025two, venegas2021ndt}. These characteristics have allowed its effective implementation to identify subsurface defects in a range of materials, including polyvinyl chloride (PVC) \cite{pt_dataset, irt_depth}, fiber-reinforced composites \cite{3d_cnn, attention_unet}, polylactic acid (PLA) components \cite{yusra_taguchi, yusra_3}, and even civil infrastructure such as concrete \cite{construction_ndt, construction_uav}.

IRT has emerged as a critical tool within NDT for identifying surface and subsurface defects by analyzing thermal behavior. IRT operates on the principle of monitoring heat propagation across the surface of a material, with deviations in thermal patterns revealing potential internal anomalies \cite{Carosena2004}. In particular, Active IRT (AIRT) enhances sensitivity to defect detection by introducing external thermal stimulation, such as flash lamps, halogen heaters, or lasers, thereby amplifying the thermal contrast between defective and intact regions \cite{AIRT2015}. This makes it especially suitable for inspecting non-metallic or multilayered materials where conventional methods may struggle. Compared to traditional subsurface detection techniques such as ultrasonic testing, eddy current testing, or radiographic imaging, AIRT offers several advantages: it is non-contact, faster, and requires minimal setup \cite{DUA2021}. These properties have enabled its wide adoption across sectors such as aerospace, energy, and civil infrastructure, where large-scale, on-site inspection is often required. In aerospace, for example, AIRT is used to identify delaminations, impact damage, and disbonds in composite components, supporting early fault detection and reducing maintenance costs \cite{TOWS2020,ma2025}.

While AIRT is characterized by being a non-contact and fast inspection technique, a challenge in AIRT sequences is the high levels of noise and low defect contrast. To enhance defect clarity and suppress noise, thermographic analysis employs data processing techniques such as Principal Component Analysis (PCA) to improve defect visualization \cite{pca_2, pct_gaussian, pca_3}. However, PCA is inherently linear and fails to capture non-linear patterns in thermographic sequences. Thus, AIRT autoencoders (AEs), such as denoising autoencoder (DAT) \cite{dat} and 1D denoising convolutional autoencoder (1D-DCAE-AIRT) \cite{1d_cnn}, have been proposed to capture non-linearities in thermographic data. Still, the aforementioned AEs suffer from two drawbacks. First, AIRT AEs are limited by the receptive fields of convolutional kernels, which restrict their ability to capture long-range dependencies. Second, these AEs require online training on all pixel thermal responses to sufficiently learn the underlying thermographic features, leading to heavy computational demands. This instigates a demand for AEs that are able to model local and global dependencies within thermographic data in a fast and efficient autoencoding strategy.

Inspired by the aforementioned factors, this paper presents an innovative deep learning framework that combines convolutional and attention-based mechanisms within an autoencoder architecture. 
The proposed masked CNN-Attention autoencoder, namely, AIRT-Masked-CAAE, is specifically designed to model the nonlinear temporal dynamics of pixel-level thermal responses while capturing long-range dependencies through a self-attention mechanism. This enables the model to focus on defect-relevant features while suppressing redundant thermal background noise. To efficiently train the introduced AIRT-Masked-CAAE, we propose a masked feature encoding strategy enabling the model to learn robust thermographic representations, while significantly reducing training time. The proposed method is evaluated on CFRP, PLA, and PVC specimens and compared against conventional thermographic autoencoding techniques, including principal component analysis (PCA), thermal signal reconstruction (TSR), and pulse-phase thermography (PPT). Experimental results demonstrate that the AIRT-Masked-CAAE improves defect visibility and reduces noise, outperforming these traditional approaches in both qualitative and quantitative evaluations.


 The key contributions of this paper are as follows:
\begin{enumerate}
    \item 
	We propose a masked CNN-Attention autoencoder, AIRT-Masked-CAAE, that simultaneously captures non-linear local and global contexts in thermographic data.
    \item 	We introduce a generalizable masked feature autoencoding strategy to enable efficient and fast training on AIRT sequences.
    \item 	We benchmark the proposed method against state-of-the-art AIRT autoencoders. Results show that the masked CNN-Attention autoencoder outperforms the state-of-the-art while reducing training time by a factor of $30\times$.
    \item 	We make our codebase publicly available to support reproducibility and further research.
\end{enumerate}


The remainder of this article is organized as follows. Section \ref{sec: relatedwork} provides an overview of the relevant literature, outlining existing approaches, their contributions, and the limitations that motivate the development of our proposed framework. Section \ref{sec: Preliminaries} introduces the necessary preliminaries, including a brief overview of active infrared thermography and autoencoders. Section \ref{sec: Methodology} details the proposed methodology, describing the masked CNN-attention autoencoder architecture, the masked feature autoencoding strategy, and the details of the implementation of training. Section \ref{sec: Experiments} presents the experimental setup, followed by a comprehensive evaluation using signal enhancement and learning-based metrics, along with benchmarking against state-of-the-art methods. The paper is concluded in \ref{sec: Conclusions}, where a summary of the findings is provided along with directions for future research.

\subsection{Related Work}
\label{sec: relatedwork}

AIRT has established itself as a key NDT technique for detecting hidden subsurface defects across a range of industrial components. Its applications have expanded rapidly in aerospace \cite{cfrp_deep}, construction \cite{thermosense_concrete}, and artwork inspection \cite{thermosense_artwork}. In the biomedical domain, thermographic datasets have also been developed for diagnostic purposes, such as arthritis classification using CNNs on knee thermograms, highlighting the cross-disciplinary adaptability of thermal imaging combined with AI \cite{bardhan2022designing}. More recently, AIRT has witnessed the adoption of AI methodologies in experimental setups to further enhance its reliability and detection accuracy. As a result, applications now span metallic crack detection via flying-spot thermography \cite{helvig2025automated}, microporous defect segmentation in composite films \cite{zhang2025automatic}, and industrial crack inspection through inductive thermography \cite{oswald2025inductive}.

The growing demand for AIRT has led to a series of investigations into neural network architectures, triggering the emergence of advanced learning-based defect detection models. To illustrate, research has fine-tuned Faster R-CNN and YOLOv5 networks for IRT defect detection \cite{flexible_framework, irt_depth}, while ConvLSTM networks have also been proposed to capture temporal dependencies for enhanced detection \cite{cnnlstm}. Beyond deep learning, Alhammad et al. (2024) \cite{alhammad2024multi} addressed this challenge in composite materials by developing Random Forest–based multi-label classification frameworks, showing that statistical features derived from thermal images outperform raw pixel data for accurate classification.

In addition to detection, various neural network architectures have been proposed for segmenting AIRT defects. U-Net variants have been applied not only in composites \cite{zhang2025automatic} but also in forged parts \cite{mueller2022defect}, while ConvLSTM networks have been used to reconstruct 3D defect depth profiles, demonstrating the benefit of combining temporal and spatial information. Fang et al. \cite{unet_study} experimentally compared state-of-the-art segmentation architectures, such as U-Net and ResNet, highlighting their effectiveness for subsurface defect segmentation. Based on these efforts, U-Net \cite{attention_unet} was proposed to further improve segmentation compared to the traditional U-Net. Despite these advances, most of these networks employ spatial filters without considering the inherent temporal features in thermographic sequences. To address this, 3D CNNs have been introduced to incorporate temporal dynamics and enhance the segmentation of subsurface defects in industrial components \cite{3d_cnn}. However, these models remain fully convolutional, and their performance is constrained by the limited receptive field of CNN layers. Consequently, attention mechanisms have been incorporated into IRT segmentation networks to overcome this limitation and capture long-range dependencies in thermographic data \cite{thermosense_attention}.

AIRT dimensionality reduction techniques form the foundation of these approaches, enabling the creation of compact thermographic representations that are then used as inputs for deep neural networks. Some of the most common methods include Thermal Signal Reconstruction (TSR) \cite{tsr, tsr_2}, Pulsed Phase Thermography (PPT) \cite{ppt_1}, and Principal Component Analysis (PCA) \cite{pct}. PCA stands out due to its effectiveness, specifically in improving the detection of hard-to-spot subsurface flaws \cite{sparse_kernel_pca}. Consequently, PCA is often used as a preprocessing step before feeding thermographic sequences into deep learning models. However, PCA and related linear methods struggle to capture the nonlinear features present in thermographic sequences. Beyond PCA, alternative physics-based approaches have been developed. For instance, Criniere et al. \cite{criniere2014inverse} compared pulsed thermography (PT) with square pulsed thermography (SPT) in CFRP-reinforced concrete, showing that SPT offers greater robustness against noise when combined with inverse thermal quadrupole models. Additionally, projected Thermal Diffusivity (PTD) analysis has been proposed for aeronautical components \cite{venegas2021ndt}, while quantitative porosity determination has been achieved through pulsed thermography coupled with XCT validation \cite{mayr2017non}. Similarly, Sun et al. \cite{sun2013analysis} analyzed a range of pulsed thermal imaging methods, including PPT, PCT, derivative, and tomography, concluding that tomography provides superior interpretability by producing spatially resolved thermal effusivity distributions. These studies underscore the enduring relevance of physics-based dimensionality reduction and characterization methods, even as deep learning gains momentum.
 


On the other hand, many recent studies have explored data-driven autoencoders for automated defect detection and dimensionality reduction. Autoencoders capture complex, non-linear features by learning efficient latent representations of input data. CNN-based autoencoders have also been employed to extract local spatial and temporal features in thermographic sequences \cite{dat, 1d_cnn, autoencoder}. For instance, 1D convolutional autoencoders have demonstrated good performance in modeling the temporal evolution of pixel responses, offering a lightweight alternative to more complex models \cite{1d_cnn}. However, due to their fully convolutional nature, they fail to detect long-range dependencies in thermographic data, crucial for identifying diffuse or irregular subsurface defects. In addition, traditional AIRT AEs require online training on all pixel thermal profiles, which tends to be time-consuming and computationally demanding. Hence, a significant gap remains in developing lightweight yet effective training strategies for models that can simultaneously extract both local and global features from thermographic sequences. Existing models often trade off between spatial resolution and temporal awareness or impose resource-intensive training schemes that limit scalability and deployment. This gap motivates a masked CNN-Attention autoencoder that integrates convolutional layers for local feature extraction with an attention mechanism to capture long-range temporal dependencies, while employing a masked feature encoding strategy to improve training efficiency and generalization.





\begin{figure}[t]
    \centering
    \includegraphics[width=0.95\textwidth]{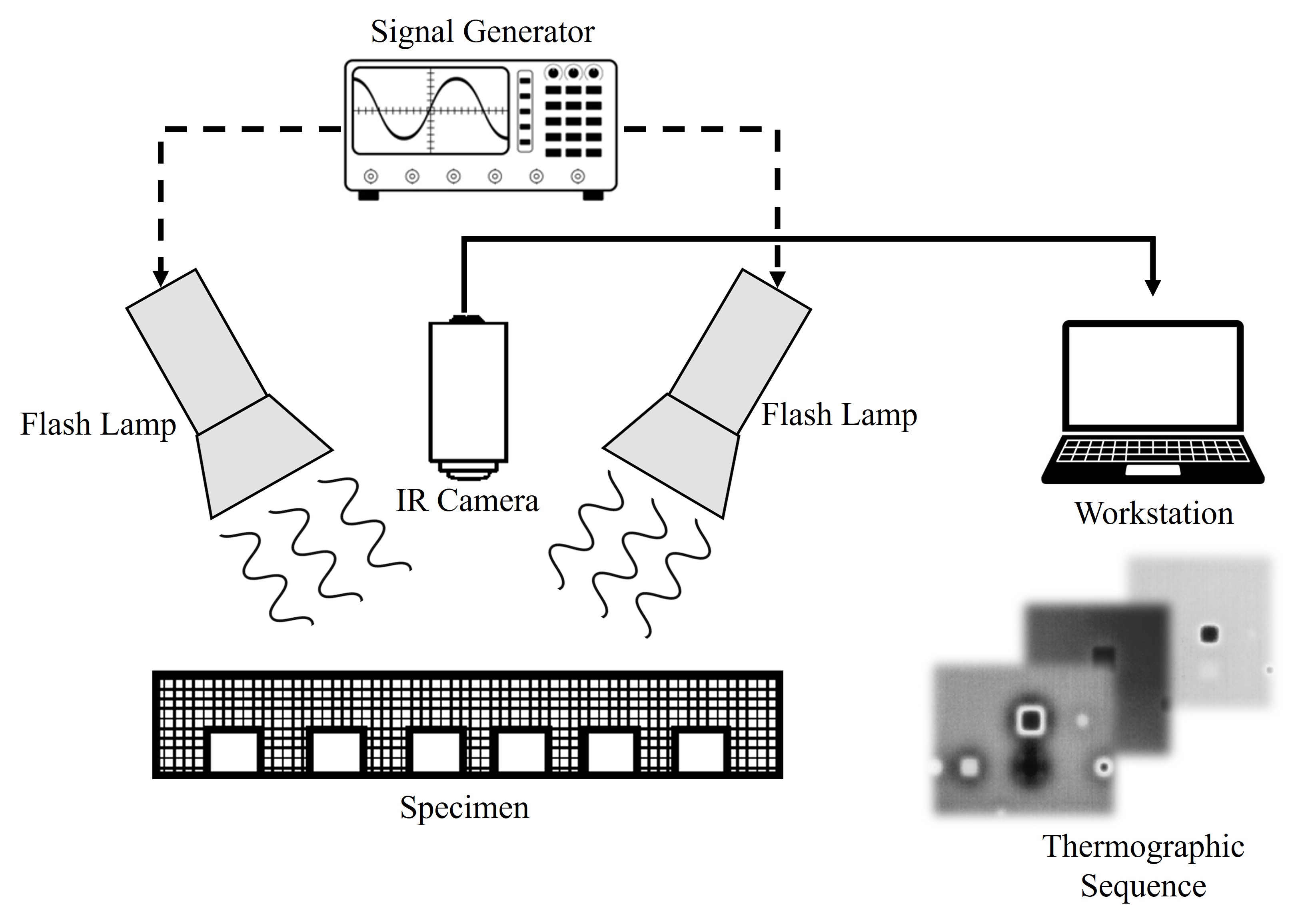}
    \caption{Typical IRT setup involving flash lamps and an IR camera. In the presence of defects, heat tends to be trapped, resulting in an abnormal thermal distribution on the specimen's surface, which is recorded by the IR camera for IRT inspection.}
    \label{fig:setup}
    \hfill
\end{figure}

\section{Preliminaries}
\label{sec: Preliminaries}
\subsection{Active Infrared Thermography}
Standard AIRT setups follow the configuration shown in Figure \ref{fig:setup}. AIRT involves a heating source, typically a halogen lamp, that excites a target specimen with a controlled heat input \cite{ndt_review}. If the specimen is sound, all pixels of the IR camera generate similar thermal profiles. Otherwise, heat is trapped, and abnormal thermal profiles are captured by the IR camera. Furthermore, thermographic inspections can be further categorized based on the type of excitation. Pulse thermography typically involves a pulse, whereas lock-in thermography involves periodic heat excitation. In both regimes, the inspection sequence is a 3D matrix, $\mathbf{S} = \{ I_{k} \}_{1}^{N_{t}}$, of shape $(N_{t}, N_{y}, N_{x})$, where $I_{k}$ is a thermogram timestamped at $k = 1, 2, \dots, N_{t}$, $N_{y}$ is the image height, and $N_{x}$ is its width. For subsequent defect analysis, $\mathbf{S}$ is reshaped to $(N_{t}, N_{y} \times N_{x})$ by a raster-like operation and centered by $\mathbf{\hat{S}} = \mathbf{S} - \mu_{k}$, where

\begin{equation}
    \mu_{k} = \frac{1}{N_{t}} \sum_{k=1}^{N_{t}} S^{(k)},
\end{equation}

\noindent and $\mathbf{\hat{S}} = \{ S^{(n)} \}_{1}^{N_{x}\times N_{y}}$ is a matrix consisting of the centered pixel-wise thermal responses.

\subsection{Autoencoders}
Prior to defect analysis in AIRT, AIRT methods reduce the dimensionality of $\mathbf{\hat{S}}$ using AEs. AEs are unsupervised neural networks commonly utilized to extract features from input data. Autoencoders compress input sequences to a latent space of lower dimensionality compared to the input. Hence, autoencoders encompass an encoder and decoder, where the encoder maps the input, $S^{(n)}$, to a latent space by

\begin{equation}
    \mathbf{z}_n = f_{\boldsymbol{\theta}}(S^{(n)}),
\end{equation}

\noindent where $f_{\boldsymbol{\theta}}(\cdot)$ is the encoder of $\theta$ weights, and $\mathbf{z}_n$ is the compressed latent vector. Similarly, the decoder regenerates the input from the latent vector by

\begin{equation}
    \tilde{S}^{(n)} = g_{\boldsymbol{\phi}}(\mathbf{z}_n),
\end{equation}

\noindent where $g_{\boldsymbol{\phi}}(\cdot)$ denotes the decoder of weights $\phi$ and $\tilde{S}^{(n)}$ is the reconstructed input. Note that $\mathbf{z}_n$ is utilized to generate the compressed latent space images for downstream defect analysis.

\begin{figure}[t]
    \centering
    \includegraphics[width=\textwidth]{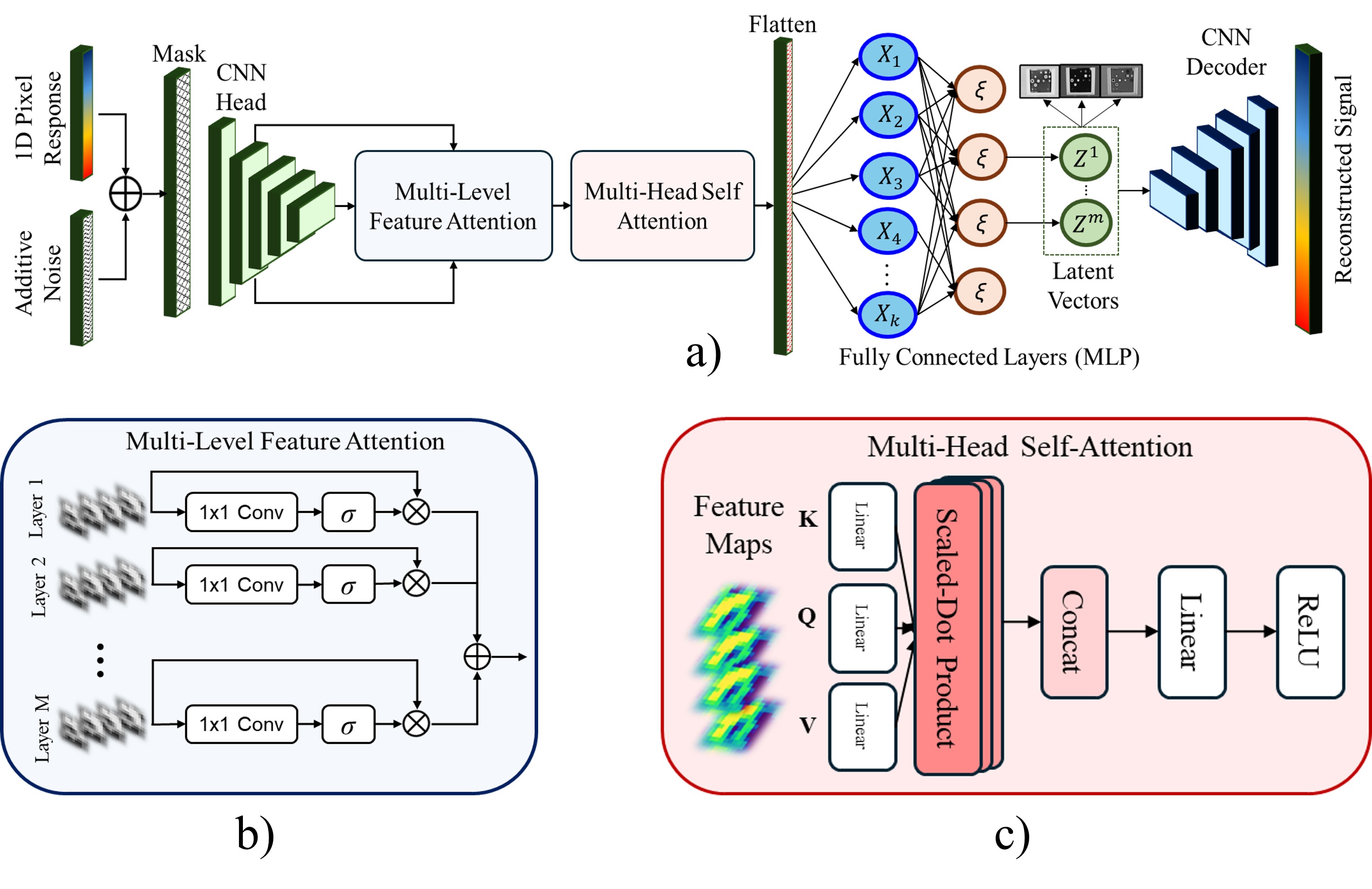}
    \caption{a) Architecture of the AIRT-Masked-CAAE. Input sequences are initially corrupted with masking and noise operations. The corrupted sequences are fed to the CNN head, followed by b) multi-level feature attention and c) multi-head attention blocks. The encoded representation is used to generate a compact latent space for downstream defect analysis tasks.}
    \label{fig:network}
    \hfill
\end{figure}

\section{Methodology}
\label{sec: Methodology}
\subsection{Network Architecture}
The architecture of the proposed AIRT-Masked-CAAE network is shown in Figure \ref{fig:network}. To ensure that the network avoids learning trivial, identity reconstruction, and focuses on input signal features, $S^{(n)}$ is subjected to a binary masking operation with additive Gaussian noise, yielding a corrupted sequence, $\hat{S}^{(n)}$. Consequently, $\hat{S}^{(n)}$ is passed through a CNN head with $L$ stacked convolutional layers to extract hierarchical local features as
\begin{equation}
    F^{(l)} = ReLU \big( W^{(l)} * F^{(l-1)} + b^{(l)} \big), 
    \quad l = 1, \dots, L,
\end{equation}
\noindent where $F^{(l)}$ represents the feature map 
obtained at the $l$-th convolutional layer, $W^{(l)}$ and $b^{(l)}$ are the learnable convolutional kernel and bias of the $l$-th layer, respectively, and $*$ denotes the convolution operation. The network CNN head generates a hierarchical set of feature maps $\{F^{(1)}, F^{(2)}, \dots, F^{(L)}\}$, each encoding localized information at different levels of abstraction, and is fed to the multi-level feature attention block.

The purpose of the multi-level feature attention block is to emphasize the most discriminative 
thermal feature maps while suppressing redundant responses. The aforementioned module is governed by learnable attention maps that adaptively weight the features extracted from different CNN layers. For each feature map 
$F^{(m)} \in \mathbb{R}^{C_m \times H \times W}$, an attention weight map is generated using a $1 \times 1$ convolution followed by a sigmoid activation as

\begin{equation}
    \alpha_m = \sigma \big( W_m^{(1 \times 1)} * F^{(m)} + b_m \big),
\end{equation}

\noindent where $W_m^{(1 \times 1)}$ and $b_m$ are the learnable $1\times 1$ kernel and $\sigma(\cdot)$ is the sigmoid activation ensuring $\alpha_m$ is bounded by $[0,1]$. The final attended representation is then obtained as

\begin{equation}
    F_{\text{att}} = \sum_{m=1}^{L} 
    \big( \alpha_m \odot F^{(m)} \big),
\end{equation}

\noindent where $\odot$ denotes element-wise multiplication. The intuition behind the resulting $F_{\text{att}}$ is that it provides a unified representation where the contribution of the feature maps from the convolutional layers of the CNN head is modulated by its learned attention map. As a result, defect-relevant information across all levels is preserved while suppressing background redundancy, thereby improving defect visualization.

Up until this stage, the network primarily focuses on extracting local features using convolutional operations and multi-level feature attention. While these mechanisms capture a hierarchical set of thermal features, long-range dependencies within input thermographic sequences are not yet captured due to the limited receptive field of convolutional layers. In AIRT, such long-range relationships are critical, as defect patterns may span temporally distant evolutions. Thus, the  multi-head self-attention module is employed to $F_{\text{att}}$ by projecting it into query, key, and value representations as

\begin{equation}
    Q = F_{\text{att}} W_Q, \quad 
    K = F_{\text{att}} W_K, \quad 
    V = F_{\text{att}} W_V,
\end{equation}

\noindent where $W_Q, W_K, W_V \in \mathbb{R}^{d \times d_k}$ are learnable  projection matrices, $d$ is the input feature dimension, and $d_k$  is the dimension of each attention head.  For the $i$-th head, attention is computed by

\begin{equation}
    \text{head}_i = 
    \text{softmax}\!\left( \frac{Q_i K_i^\top}{\sqrt{d_k}} \right) V_i,
    \quad i = 1, \dots, H,
\end{equation}

\noindent where $H$ is the number of heads. The outputs from all heads are then concatenated and passed through a 
linear transformation followed by a ReLU activation as

\begin{equation}
    F_{\text{MHA}} = 
    \text{ReLU}\!\Big( 
        \text{Concat}(\text{head}_1, \dots, \text{head}_H) W_O + b_O
    \Big),
\end{equation}

\noindent where $W_O \in \mathbb{R}^{Hd_k \times d}$ and $b_O$ are the learnable parameters of the output projection.

The feature maps produced by the multi-head self-attention block are passed to a multi-layer perceptron (MLP), which compresses them into a compact latent representation $\mathbf{z}_n$. Note that $\mathbf{z}_n$ is utilized to formulate the latent images for subsequent defect analysis. Finally, the latent representation $\mathbf{z}_n$ is then decoded through a sequence of convolutional layers to reconstruct the thermographic input. The network is trained using the combined reconstruction-knowledge distillation loss function defined in \cite{pca_guided_ae} as

\begin{equation}
\mathcal{L}_{\text{total}} =
\mathcal{L}_{\text{rec}} +
\alpha\, \mathcal{L}_{\text{KD}},
\label{eq:total_loss}
\end{equation}

\noindent where 

\begin{equation}
\mathcal{L}_{\text{rec}} =
\frac{1}{N} \sum_{i=1}^{N} \| \tilde{S}_{i}^{(n)} - S_{i}^{(n)} \|_2^2,
\label{eq:recon_loss}
\end{equation}

\noindent and 

\begin{equation}
\mathcal{L}_{\text{KD}} =
1 -
\frac{
\langle \mathbf{z}_n,\, \mathbf{z}'_n \rangle
}{
\|\mathbf{z}_n\|_2\, \|\mathbf{z}'_n\|_2
}.
\label{eq:kd_loss}
\end{equation}

\noindent Note that $N$ is the number of pixels in the batch, $\tilde{S}_{i}^{(n)}$ denotes the reconstructed temporal response of the $i^{th}$ pixel, and $\mathbf{z}_{n}$ is the PCA latent vector as outlined in \cite{pca_guided_ae}.

\begin{algorithm}[t]
\caption{Masked Sequence Autoencoding}
\begin{algorithmic}[1]
\STATE \textbf{Inputs:} Thermographic sequences $\mathbf{S}=\{S^{(n)}\}_{n=1}^N$, noise variance $\sigma^2$, batch size $B$ \\
\STATE \textbf{Outputs:} Trained parameters, $\theta^*$ and $\phi^*$, latent vectors $\{\mathbf{z}_{n}\}$
\WHILE{not converged}
  \STATE \textbf{Sample} batch $\mathcal{B}$ of size $B$
    \STATE \textbf{Generate} random binary mask $M^{(n)}$
    \STATE \textbf{Corrupt} input $\tilde{S}^{(n)} \leftarrow M^{(n)} \odot S^{(n)} + \mathcal{N}(0,\sigma^2)$
    \STATE \textbf{Forward:}
    \STATE \hspace{0.8em} $F^{(1{:}L)} \leftarrow$ \textit{CNN Head}$\big(\tilde{S}^{(n)}\big)$
    \STATE \hspace{0.8em} $F_{\text{att}} \leftarrow$ \textit{Multi-Level Feature Attention}$\big(F^{(1{:}L)}\big)$
    \STATE \hspace{0.8em} $F_{\text{MHA}} \leftarrow$ \textit{Multi-Head Self-Attention}$\big(F_{\text{att}}\big)$
    \STATE \hspace{0.8em} $\mathbf{z}_{n} \leftarrow$ \textit{MLP}$\big(F_{\text{MHA}}\big)$
    \STATE \hspace{0.8em} $\hat{S}^{(n)} \leftarrow$ \textit{Decoder}$\big(\mathbf{z}_{n}\big)$
  \STATE \textbf{Compute} training loss $\mathcal{L} = \mathcal{L}_{\text{rec}} + \mathcal{L}_{\text{KD}}$
  \STATE \textbf{Update} $\theta^* \leftarrow \theta - \eta \nabla_\theta \mathcal{L}$, $\phi^* \leftarrow \phi - \eta \nabla_\phi \mathcal{L}$
\ENDWHILE
\STATE \textbf{return} $\theta^*$, $\phi^*$, $\{\mathbf{z}_{n}\}$
\end{algorithmic}
\label{alg:training}
\end{algorithm}

\subsection{Masked Sequence Autoencoding}
Traditional AIRT autoencoders require online training of the network on all raw pixel responses $S^{(n)}$. This approach suffers from two major drawbacks. First, the network may converge to trivial identity reconstruction, leading to sub-optimal latent representations. Second, training on the entire set of pixel responses can be time and computationally intensive, especially when training deep architectures such as the AIRT-Masked-CAAE. Thus, instead of training the AIRT-Masked-CAAE on all $S^{(n)}$ samples, this work proposes a fast, generalizable training strategy by means of masked sequence autoencoding.

The masked sequence autoencoding strategy, outlined in Alg. \ref{alg:training}, involves injecting the input with zero mean additive Gaussian noise and masking random patches of the input before being passed through the network. Accordingly, the corrupted sequence, $\hat{S}_{i}^{(n)}$ is formulated by

\begin{equation}
    \hat{S}^{(n)} = M \odot S^{(n)} + \mathcal{N}(0,\sigma^2),
\end{equation}

\noindent where $M$ is a 1-D binary mask indicating visible ($1$) and masked ($0$) patches, 
and $\mathcal{N}(0,\sigma^2)$ represents additive Gaussian noise with zero mean and variance $\sigma^2$. Note that due to the masking operation, only a subset of the thermographic data is used at each training step, speeding up training by a factor of $30\times$. The masking operation enables the model to focus on extracting robust and generalizable representations rather than trivial pixel-level reconstructions without requiring access to all samples of pixel responses. Accordingly, the corrupted input $\hat{S}^{(n)}$ is fed through the encoder, attention modules, and decoder, with the objective of reconstructing the original sequence $S^{(n)}$. On the other hand, while the input is partially masked and perturbed with noise, the training loss defined in Eq. \ref{eq:total_loss} remains unchanged and is computed with respect to the original, uncorrupted sequence.

After developing the AIRT-Masked-CAAE architecture and its masked sequence autoencoding strategy, Bayesian optimization is carried out to select the optimal hyperparameters of the network. The AIRT-Masked-CAAE hyperparameters are the number of convolutional layers in the CNN head and their kernel sizes, the number of heads in the self-attention block, the number of fully connected layers in the MLP bottleneck and its latent vector size, and the number of training epochs. Accordingly, the CNN head comprises 3 convolutional layers with $3\times 3$ kernels. The multi-head self-attention block consists of four attention heads. The MLP generates a latent vector of size 32; i.e., each inspection sequence is compressed to 32 images, which are utilized for subsequent defect analysis. Finally, the network is trained using ADAM optimizer, with a learning rate of $2e^{-5}$, and a batch size of 128.

\begin{figure}[t]
    \centering
    \includegraphics[width=\textwidth]{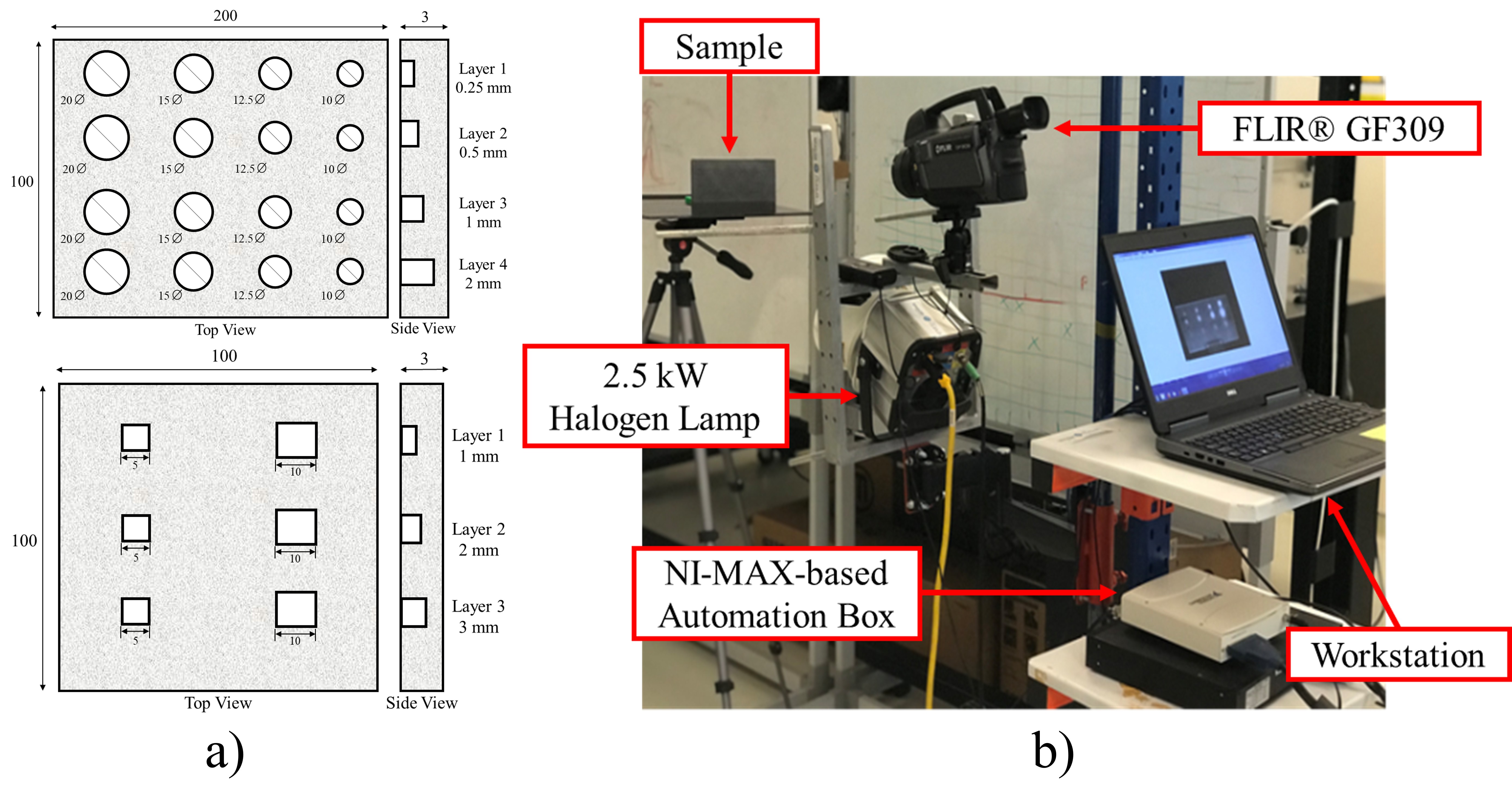}
    \caption{a) Data collection setup for testing the AIRT-Masked-CAAE. b) CFRP and PLA samples to validate the AIRT-Masked-CAAE. The dimensions are expressed in mm.}
    \label{fig:exp_setup}
    \hfill
\end{figure}

\section{Experiments}
\label{sec: Experiments}
\subsection{Experimental Setup}
The proposed AIRT-Masked-CAAE is rigorously evaluated on three panels: one of CFRP and two of PLA material. The panels were produced using 3D printing with Teflon insertions to simulate real-world defects. Samples 1 and 2 are generated with 16 inserts, while sample 3 is created with 6, see Figure \ref{fig:exp_setup}a. All defects are distributed at depth layers of 0.25, 0.50, 1.0, 2.0 mm. Consequently, AIRT inspection sequences are created using the data collection setup in Figure \ref{fig:exp_setup}b, where samples are heated by a 2.5 kW halogen lamp controlled by an NI-MAX-based automation box. Three pulse durations are employed to investigate the capability of the AIRT-Masked-CAAE to capture thermal features under different excitations. Finally, GF309 from FLIR® featuring an Indium Antimonide InSb cooled thermal imager is employed to capture the inspection sequences. The thermal camera is placed at 10 cm from the panel, while the halogen lamps are at 8 cm and 15$^\circ$ incidence angle. As such, non-uniform heating is intentionally introduced for evaluation purposes. 

\begin{table*}[p]
\centering
\renewcommand{\arraystretch}{2}
\caption{Qualitative comparisons between state-of-the-art AIRT dimensionality reduction techniques: TSR, PCA, DAT \cite{dat}, 1D-DCAE-AIRT \cite{1d_cnn}, and C-AET \cite{constrained_ae} against the proposed CNN-Attention autoencoder.}
\resizebox{\textwidth}{!}{%
\begin{tabular}{|>{\centering\arraybackslash}m{2.5cm}|>{\centering\arraybackslash}m{4cm}|>{\centering\arraybackslash}m{4cm}|>{\centering\arraybackslash}m{4cm}|}
\hline
\textbf{Method} & \textbf{CFRP} & \textbf{PLA} & \textbf{PVC} \\ \hline

\textbf{Raw} &
{\vspace{3pt}\includegraphics[width=3.75cm, height=2.5cm]{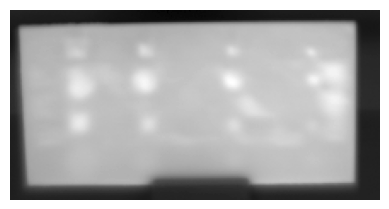}} &
{\vspace{3pt}\includegraphics[width=3.75cm, height=2.5cm]{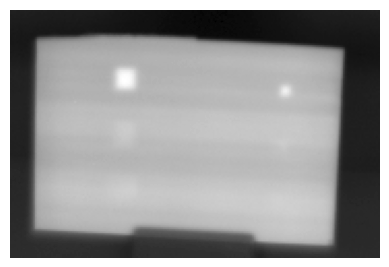}} &
{\vspace{3pt}\includegraphics[width=3.25cm, height=2.5cm]{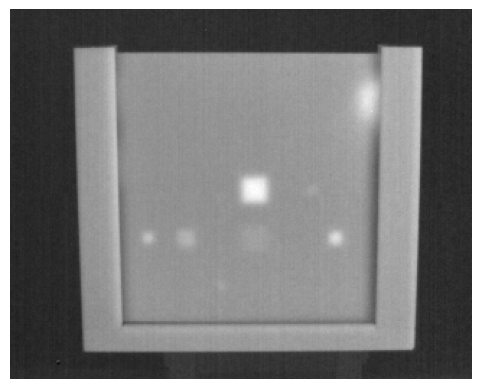}} \\ \hline

\textbf{TSR} &
{\vspace{3pt}\includegraphics[width=3.75cm, height=2.5cm]{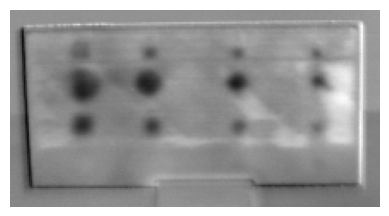}} &
{\vspace{3pt}\includegraphics[width=3.75cm, height=2.5cm]{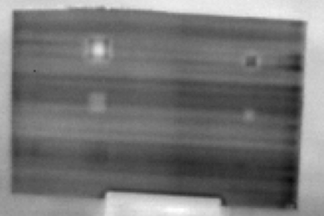}} &
{\vspace{3pt}\includegraphics[width=3.25cm, height=2.5cm]{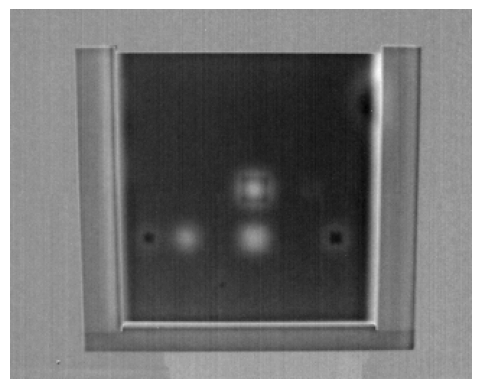}} \\ \hline

\textbf{PCA} &
{\vspace{3pt}\includegraphics[width=3.75cm, height=2.5cm]{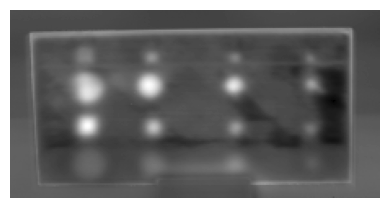}} &
{\vspace{3pt}\includegraphics[width=3.75cm, height=2.5cm]{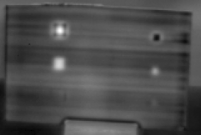}} &
{\vspace{3pt}\includegraphics[width=3.25cm, height=2.5cm]{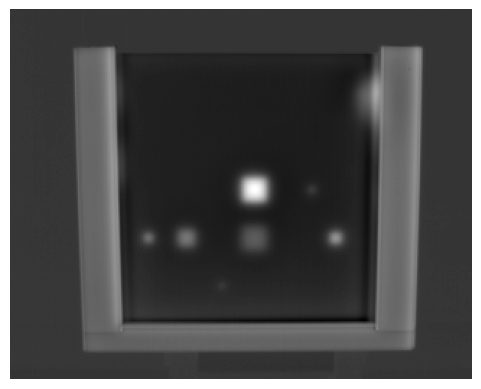}} \\ \hline

\textbf{DAT} &
{\vspace{3pt}\includegraphics[width=3.75cm, height=2.5cm]{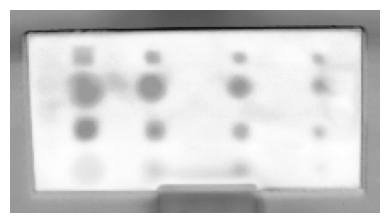}} &
{\vspace{3pt}\includegraphics[width=3.75cm, height=2.5cm]{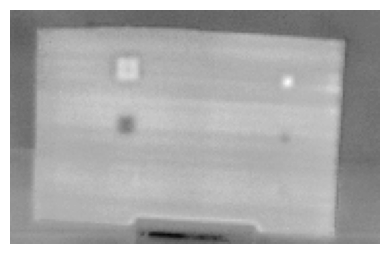}} &
{\vspace{3pt}\includegraphics[width=3.25cm, height=2.5cm]{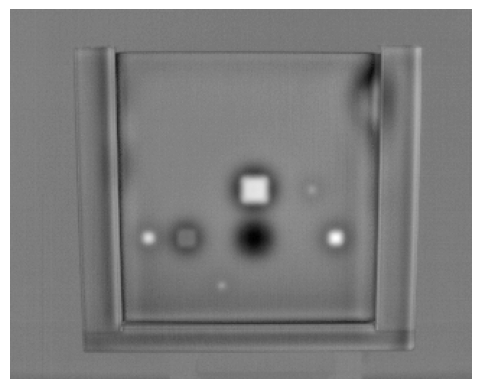}} \\ \hline

\textbf{1D-DCAE-AIRT} &
{\vspace{3pt}\includegraphics[width=3.75cm, height=2.5cm]{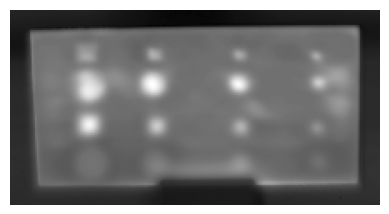}} &
{\vspace{3pt}\includegraphics[width=3.75cm, height=2.5cm]{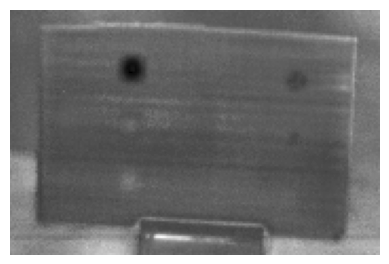}} &
{\vspace{3pt}\includegraphics[width=3.25cm, height=2.5cm]{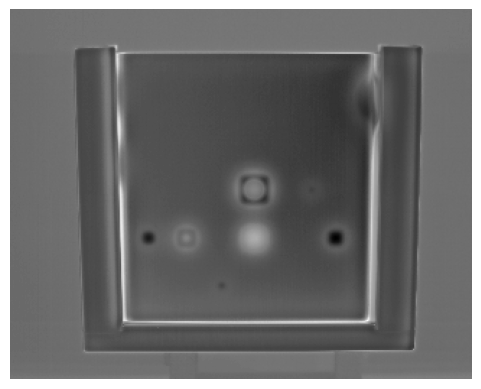}} \\ \hline

\textbf{C-AET} &
{\vspace{3pt}\includegraphics[width=3.75cm, height=2.5cm]{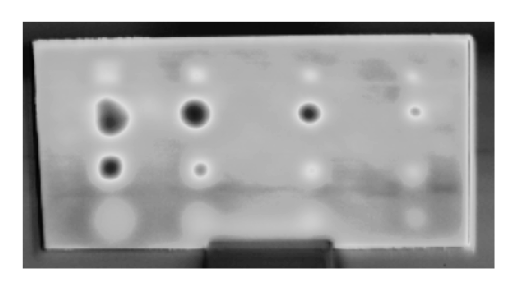}} &
{\vspace{3pt}\includegraphics[width=3.75cm, height=2.5cm]{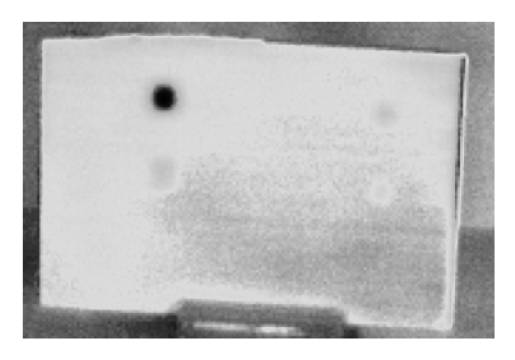}} &
{\vspace{3pt}\includegraphics[width=3.25cm, height=2.5cm]{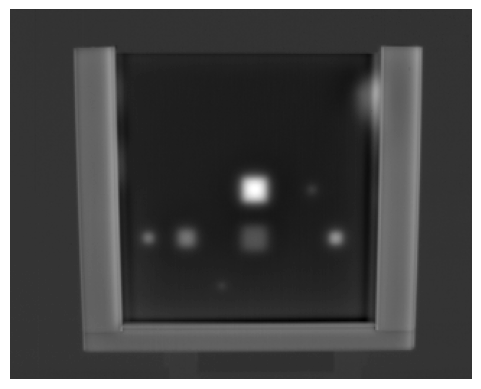}} \\ \hline

\textbf{Ours} &
{\vspace{3pt}\includegraphics[width=3.75cm, height=2.5cm]{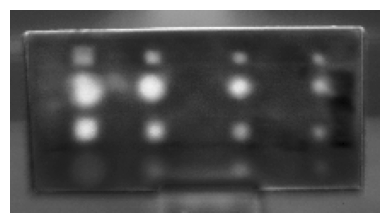}} &
{\vspace{3pt}\includegraphics[width=3.75cm, height=2.5cm]{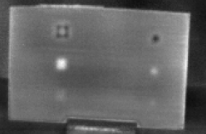}} &
{\vspace{3pt}\includegraphics[width=3.25cm, height=2.5cm]{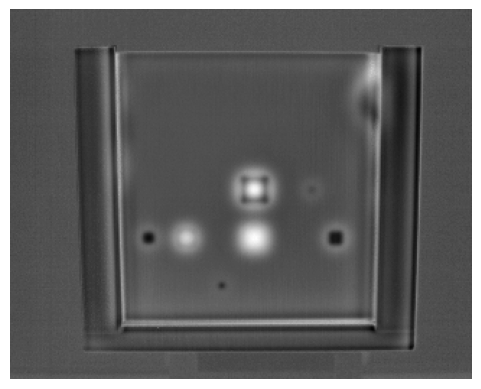}} \\ \hline

\end{tabular}
}
\label{table:qual_comparison}
\end{table*}

In addition to the CFRP and PLA panels, the AIRT-Masked-CAEE is evaluated on the IRT-PVC dataset \cite{irt_depth}. This allows for additional evaluation of the framework on PVC panels and evaluation of the network's latent representations for neural network-based evaluation, as highlighted in \cite{pca_guided_ae}. The IRT-PVC dataset contains 38 inspected samples with back-drilled holes at depths ranging from 2.5 to 4.5 mm. Having the CFRP, PLA, and PVC inspection sequences, the performance of the AIRT-Masked-CAAE is validated based on defect visibility and signal enhancement, discussed in Section \ref{sec:signal}, and neural network-based evaluation, where the AE latent images are used as an input to a segmentation network, discussed in Section \ref{sec:nn}. Accordingly, the metrics for the first evaluation routine that quantify signal enhancement are contrast and signal-to-noise ratio (SNR), defined as

\begin{equation}
    \text{Contrast} = 
    \frac{\left| \left(\frac{1}{N}\sum_{p=1}^{N} Y_{d}(p)\right) - \left(\frac{1}{M}\sum_{q=1}^{M} Y_{s}(q)\right) \right|}
    {\left(\frac{1}{N}\sum_{p=1}^{N} Y_{d}(p)\right) + \left(\frac{1}{M}\sum_{q=1}^{M} Y_{s}(q)\right)},
\end{equation}

\begin{equation}
    \text{SNR} = 
    \frac{\left| \left(\frac{1}{N}\sum_{p=1}^{N} Y_{d}(p)\right) - \left(\frac{1}{M}\sum_{q=1}^{M} Y_{s}(q)\right) \right|}
    {\sigma_{s}},
\end{equation}

\noindent where $N$ denotes the total number of pixels in the defective region $Y_{d}$, with $Y_{d}(p)$ representing the $p^{th}$ pixel intensity in that region. $M$ refers to the number of pixels in the sound region $Y_{s}$, with $Y_{s}(q)$ being the $q^{th}$ pixel intensity, while $\sigma_{s}$ corresponds to the standard deviation of pixel values in the sound region $Y_{s}$. For the segmentation-based validation, the evaluation metric is the intersection over union (IoU) formulated as

\begin{equation}
    \text{IoU} = \frac{|P \cap G|}{|P \cup G|},
\end{equation}

\noindent where $P$ and $G$ represent the predicted and ground-truth segmentation masks. It is also worth mentioning that both evaluation routines in Sections \ref{sec:signal} and \ref{sec:nn} present benchmarks of the proposed framework against state-of-the-art AIRT dimensionality reduction methods and autoencoders. Section \ref{subsection:denoising} evaluates the denoising performance of the AIRT-Masked-CAAE, while Section \ref{subsection:ablation} presents ablation studies.

\begin{table}[t]
\centering
\caption{Contrast and SNR for the AIRT-Masked-CAAE benchmarked against state-of-the-art dimensionality reduction methods on CFRP samples.}
\label{tab:signal_cfrp}
\resizebox{\textwidth}{!}{%
\begin{tabular}{|c|cc|cc|cc|cc|cc|}
\hline
\textbf{Defect Class} & \multicolumn{2}{c|}{\textbf{0.25 mm}}                      & \multicolumn{2}{c|}{\textbf{0.5 mm}}                       & \multicolumn{2}{c|}{\textbf{1.0 mm}}                       & \multicolumn{2}{c|}{\textbf{2.0 mm}}                       & \multicolumn{2}{c|}{\textbf{All}}                                \\ \hline
\textbf{Metric}       & \multicolumn{1}{c|}{\textbf{Contrast}} & \textbf{SNR (dB)} & \multicolumn{1}{c|}{\textbf{Contrast}} & \textbf{SNR (dB)} & \multicolumn{1}{c|}{\textbf{Contrast}} & \textbf{SNR (dB)} & \multicolumn{1}{c|}{\textbf{Contrast}} & \textbf{SNR (dB)} & \multicolumn{1}{c|}{\textbf{Contrast}}    & \textbf{SNR (dB)}    \\ \hline
Raw                   & \multicolumn{1}{c|}{0.202}             & 19.97             & \multicolumn{1}{c|}{0.246}             & 23.79             & \multicolumn{1}{c|}{0.218}             & 24.91             & \multicolumn{1}{c|}{0.179}             & 22.66             & \multicolumn{1}{c|}{0.221}                & 22.83                \\ \hline
TSR                   & \multicolumn{1}{c|}{0.319}             & 26.84             & \multicolumn{1}{c|}{0.278}             & 26.78             & \multicolumn{1}{c|}{0.321}             & 27.97             & \multicolumn{1}{c|}{0.299}             & 22.13             & \multicolumn{1}{c|}{0.305}                & 25.93                \\ \hline
PCA                   & \multicolumn{1}{c|}{0.283}             & 30.89             & \multicolumn{1}{c|}{0.307}             & 23.81             & \multicolumn{1}{c|}{0.296}             & 26.83             & \multicolumn{1}{c|}{0.361}             & 39.74             & \multicolumn{1}{c|}{0.311}                & 30.32                \\ \hline
DAT                   & \multicolumn{1}{c|}{0.326}             & 30.11             & \multicolumn{1}{c|}{0.433}             & 34.89             & \multicolumn{1}{c|}{0.437}             & 37.61             & \multicolumn{1}{c|}{0.266}             & 27.14             & \multicolumn{1}{c|}{0.366}                & 32.44                \\ \hline
1D-DCAE-AIRT          & \multicolumn{1}{c|}{0.386}             & 32.04             & \multicolumn{1}{c|}{0.407}             & 36.66             & \multicolumn{1}{c|}{0.423}             & 32.26             & \multicolumn{1}{c|}{0.347}             & 29.91             & \multicolumn{1}{c|}{0.391}                & 32.71                \\ \hline
C-AET                 & \multicolumn{1}{c|}{0.319}             & 28.21             & \multicolumn{1}{c|}{0.346}             & 38.02             & \multicolumn{1}{c|}{0.372}             & 33.44             & \multicolumn{1}{c|}{0.318}             & 30.01             & \multicolumn{1}{c|}{0.339}                & 32.42                \\ \hline
{\textbf{Ours}}   & \multicolumn{1}{c|}{0.707}             & 44.23             & \multicolumn{1}{c|}{0.721}             & 47.11             & \multicolumn{1}{c|}{0.776}             & 51.83             & \multicolumn{1}{c|}{0.618}             & 37.87             & \multicolumn{1}{c|}{{\textbf{0.706}}} & {\textbf{45.26}} \\ \hline
\end{tabular}%
}
\end{table}

\begin{table}[t]
\centering
\caption{Contrast and SNR for the AIRT-Masked-CAAE benchmarked against state-of-the-art dimensionality reduction methods on CFRP samples.}
\label{tab:signal_pla}
\resizebox{\textwidth}{!}{%
\begin{tabular}{|c|cc|cc|cc|cc|}
\hline
\textbf{Defect Class} & \multicolumn{2}{c|}{\textbf{1.0 mm}}                       & \multicolumn{2}{c|}{\textbf{2.0 mm}}                       & \multicolumn{2}{c|}{\textbf{3.0 mm}}                       & \multicolumn{2}{c|}{\textbf{All}}                          \\ \hline
\textbf{Metric}       & \multicolumn{1}{c|}{\textbf{Contrast}} & \textbf{SNR (dB)} & \multicolumn{1}{c|}{\textbf{Contrast}} & \textbf{SNR (dB)} & \multicolumn{1}{c|}{\textbf{Contrast}} & \textbf{SNR (dB)} & \multicolumn{1}{c|}{\textbf{Contrast}} & \textbf{SNR (dB)} \\ \hline
Raw                   & \multicolumn{1}{c|}{0.312}             & 24.67             & \multicolumn{1}{c|}{0.297}             & 22.40             & \multicolumn{1}{c|}{0.281}             & 21.40             & \multicolumn{1}{c|}{0.296}             & 22.83             \\ \hline
TSR                   & \multicolumn{1}{c|}{0.369}             & 28.76             & \multicolumn{1}{c|}{0.134}             & 25.97             & \multicolumn{1}{c|}{0.113}             & 23.10             & \multicolumn{1}{c|}{0.205}             & 25.93             \\ \hline
PCA                   & \multicolumn{1}{c|}{0.371}             & 33.63             & \multicolumn{1}{c|}{0.252}             & 29.64             & \multicolumn{1}{c|}{0.219}             & 27.69             & \multicolumn{1}{c|}{0.281}             & 30.32             \\ \hline
DAT                   & \multicolumn{1}{c|}{0.469}             & 35.67             & \multicolumn{1}{c|}{0.438}             & 33.33             & \multicolumn{1}{c|}{0.371}             & 28.34             & \multicolumn{1}{c|}{0.426}             & 32.44             \\ \hline
1D-DCAE-AIRT          & \multicolumn{1}{c|}{0.479}             & 35.07             & \multicolumn{1}{c|}{0.425}             & 33.01             & \multicolumn{1}{c|}{0.389}             & 30.07             & \multicolumn{1}{c|}{0.431}             & 32.71             \\ \hline
C-AET                 & \multicolumn{1}{c|}{0.448}             & 35.46             & \multicolumn{1}{c|}{0.393}             & 30.30             & \multicolumn{1}{c|}{0.369}             & 28.36             & \multicolumn{1}{c|}{0.403}             & 31.37             \\ \hline
\textbf{Ours}         & \multicolumn{1}{c|}{0.751}             & 48.54             & \multicolumn{1}{c|}{0.691}             & 44.19             & \multicolumn{1}{c|}{0.483}             & 38.86             & \multicolumn{1}{c|}{\textbf{0.641}}    & \textbf{43.86}    \\ \hline
\end{tabular}%
}
\end{table}

\subsection{Signal Enhancement Evaluation} \label{sec:signal}
Dimensionality reduction techniques in AIRT offer the advantage of reducing data dimensionality while enhancing the visibility of defects against the background. Improved defect visibility tends to be correlated with improved downstream defect analysis tasks. Thus, the signal enhancement evaluation presented in this section evaluates the efficacy of the AIRT-Masked-CAAE in enhancing defect visibility while suppressing background noise in terms of contrast and SNR. Table \ref{table:qual_comparison} visualizes the compressed representations of CFRP, PLA, and PVC panels using our AIRT-Masked-CAAE and compares it to PCA \cite{pca_2}, TSR \cite{tsr}, DAT \cite{dat}, 1D-DCAE-AIRT \cite{1d_cnn}, and C-AET \cite{constrained_ae}. On the other hand, tables \ref{tab:signal_cfrp}, \ref{tab:signal_pla}, and \ref{tab:signal_pvc} benchmarks and quantify signal enhancement obtained by the proposed framework on the CFRP, PLA, and PVC specimens and benchmark it against TSR, PCA, DAT, 1D-DCAE-AIRT, and C-AET. Note that the CFRP samples have defects at depths of 0.25, 0.5, 1.0, and 2.0 mm. The PLA samples exhibit defects at depths of 1.0, 2.0, and 3.0 mm, whereas the PVC specimens contain defects at depths of 2.5, 3.0, 3.5, 4.0, and 4.5 mm. On the other hand, in contrast to the aforementioned methods.

The results in Tables \ref{tab:signal_cfrp}, \ref{tab:signal_pla}, and \ref{tab:signal_pvc} show that the defect signal tends to increase with decreasing defect depth across all samples. This is expected since shallower defects exhibit stronger heat diffusion responses compared to deeper defects. Nevertheless, the significant contrasts and SNR obtained for all defects are of the same order of magnitude. For instance, the proposed AIRT-Masked-CAAE increases the contrast by approximately 50\% and SNR by 20 db compared to the raw thermograms. On the other hand, the proposed framework outperforms traditional and learning-based AIRT dimensionality reduction methods. According to Table \ref{table:qual_comparison}, the visualizations highlight sharper defect boundaries, reduced halo artifacts, and superior suppression of background weave and non-uniform heating effects. Quantitatively, contrast improvements of up to 25\% and SNR gains exceeding 10 dB are observed compared to the strongest baseline, such as 1D-DCAE-AIRT. These improvements stem from the masked feature autoencoding strategy, which prevents trivial identity reconstruction and compels the network to focus on defect-relevant cues, the multi-level feature attention that amplifies salient channels across convolutional layers, and the self-attention block that captures long-range spatial and temporal dependencies. It is also worth highlighting that the aforementioned methods require access to all samples during training. To the contrary, the AIRT-Masked-CAAE leverages the masked training regime, enabling fast optimization without sacrificing generalizability and highlighting the AIRT-Masked-CAAE as a powerful dimensionality reduction approach in active infrared thermography.

\begin{table}[t]
\centering
\caption{Contrast and SNR for the AIRT-Masked-CAAE benchmarked against state-of-the-art dimensionality reduction methods on PVC samples.}
\label{tab:signal_pvc}
\resizebox{\textwidth}{!}{%
\begin{tabular}{|c|cc|cc|cc|cc|cc|cc|}
\hline
\textbf{Defect Class} & \multicolumn{2}{c|}{\textbf{2.5 mm}}                       & \multicolumn{2}{c|}{\textbf{3.0 mm}}                       & \multicolumn{2}{c|}{\textbf{3.5 mm}}                       & \multicolumn{2}{c|}{\textbf{4.0 mm}}                       & \multicolumn{2}{c|}{\textbf{4.5 mm}}                       & \multicolumn{2}{c|}{\textbf{All}}                          \\ \hline
\textbf{Metric}       & \multicolumn{1}{c|}{\textbf{Contrast}} & \textbf{SNR (dB)} & \multicolumn{1}{c|}{\textbf{Contrast}} & \textbf{SNR (dB)} & \multicolumn{1}{c|}{\textbf{Contrast}} & \textbf{SNR (dB)} & \multicolumn{1}{c|}{\textbf{Contrast}} & \textbf{SNR (dB)} & \multicolumn{1}{c|}{\textbf{Contrast}} & \textbf{SNR (dB)} & \multicolumn{1}{c|}{\textbf{Contrast}} & \textbf{SNR (dB)} \\ \hline
Raw                   & \multicolumn{1}{c|}{0.285}             & 24.61             & \multicolumn{1}{c|}{0.301}             & 25.48             & \multicolumn{1}{c|}{0.315}             & 27.33             & \multicolumn{1}{c|}{0.346}             & 28.01             & \multicolumn{1}{c|}{0.374}             & 29.83             & \multicolumn{1}{c|}{0.324}             & 27.05             \\ \hline
TSR                   & \multicolumn{1}{c|}{0.455}             & 22.86             & \multicolumn{1}{c|}{0.476}             & 23.94             & \multicolumn{1}{c|}{0.512}             & 24.67             & \multicolumn{1}{c|}{0.521}             & 25.05             & \multicolumn{1}{c|}{0.561}             & 26.34             & \multicolumn{1}{c|}{0.505}             & 24.57             \\ \hline
PCA                   & \multicolumn{1}{c|}{0.335}             & 23.91             & \multicolumn{1}{c|}{0.348}             & 24.65             & \multicolumn{1}{c|}{0.389}             & 25.52             & \multicolumn{1}{c|}{0.375}             & 27.71             & \multicolumn{1}{c|}{0.458}             & 28.56             & \multicolumn{1}{c|}{0.381}             & 26.06             \\ \hline
DAT                   & \multicolumn{1}{c|}{0.491}             & 28.07             & \multicolumn{1}{c|}{0.523}             & 31.22             & \multicolumn{1}{c|}{0.508}             & 30.17             & \multicolumn{1}{c|}{0.557}             & 32.94             & \multicolumn{1}{c|}{0.550}             & 34.75             & \multicolumn{1}{c|}{0.526}             & 31.43             \\ \hline
1D-DCAE-AIRT          & \multicolumn{1}{c|}{0.502}             & 32.27             & \multicolumn{1}{c|}{0.493}             & 33.94             & \multicolumn{1}{c|}{0.537}             & 35.63             & \multicolumn{1}{c|}{0.544}             & 36.12             & \multicolumn{1}{c|}{0.580}             & 40.79             & \multicolumn{1}{c|}{0.531}             & 35.75             \\ \hline
C-AET                 & \multicolumn{1}{c|}{0.439}             & 32.05             & \multicolumn{1}{c|}{0.468}             & 34.72             & \multicolumn{1}{c|}{0.501}             & 33.96             & \multicolumn{1}{c|}{0.495}             & 36.88             & \multicolumn{1}{c|}{0.582}             & 36.90             & \multicolumn{1}{c|}{0.497}             & 34.90             \\ \hline
\textbf{Ours}         & \multicolumn{1}{c|}{0.837}             & 51.78             & \multicolumn{1}{c|}{0.846}             & 55.13             & \multicolumn{1}{c|}{0.782}             & 49.27             & \multicolumn{1}{c|}{0.821}             & 50.01             & \multicolumn{1}{c|}{0.668}             & 43.22             & \multicolumn{1}{c|}{\textbf{0.791}}    & \textbf{49.88}    \\ \hline
\end{tabular}%
}
\end{table}

\begin{table}[t]
\centering
\caption{Obtained validation and testing IoU for the IRT-PVC defect classes using a segmentation U-Net trained on AIRT-Masked-CAAE latent images.}
\label{tab:class_iou}
\resizebox{\textwidth}{!}{%
\begin{tabular}{|c|c|c|c|c|c|c|}
\hline
\textbf{Defect Class} & \textbf{2.5 mm} & \textbf{3.0 mm} & \textbf{3.5 mm} & \textbf{4.0 mm} & \textbf{4.5 mm} & \textbf{Aggregate} \\ \hline
Validation IoU        & 0.866           & 0.804           & 0.815           & 0.868           & 0.850           & 0.841              \\ \hline
Testing IoU           & 0.844           & 0.829           & 0.839           & 0.852           & 0.817           & 0.836              \\ \hline
\end{tabular}%
}
\end{table}

\begin{table*}[!ht]
\centering
\caption{AIRT-Masked-CAAE validated and benchmarked using the proposed neural network-based evaluation, AE latent representation is utilized as input to a segmentation neural network, and the IoU serves as an evaluation metric.}
\label{tab:bench_iou}
\resizebox{\textwidth}{!}{%
\begin{tabular}{|c|c|c|c|c|c|c|c|}
\hline
\textbf{Method} & \textbf{Raw} & \textbf{TSR} & \textbf{PCA} & \textbf{DAT} & \textbf{1D-DCAE-AIRT} & \textbf{C-AET} & \textbf{Ours} \\ \hline
Validation IoU  & 0.751        & 0.738        & 0.743        & 0.209        & 0.763                 & 0.523          & \textbf{0.841} \\ \hline
Testing IoU     & 0.759        & 0.734        & 0.721        & 0.221        & 0.754                 & 0.517          & \textbf{0.836} \\ \hline
\end{tabular}%
}
\end{table*}

\subsection{Neural Network-Based Evaluation} \label{sec:nn}
While the contrast and SNR metrics evaluate the efficacy of the AIRT-Masked-CAAE in enhancing defect visibility, neural network-based evaluation validates the feasibility of the AIRT-Masked-CAAE latent representation in AI-driven defect analysis tasks. This evaluation routine involves passing the latent images of the AIRT-Masked-CAAE into a U-Net segmentation network. The network mimics the U-Net tested in \cite{pt_dataset} and is trained on 26 samples, validated on 6, and tested on 6 additional samples. Accordingly, the IoU serves as the evaluation metric and is reported for the validation and testing sets. Note that improved latent representations tend to result in higher IoUs, accompanied by enhanced defect clarity.

Table \ref{tab:class_iou} outlines the validation and testing IoUs for each defect depth class ranging from 2.5 mm to 4.5 mm. The obtained IoUs highlight that the U-Net segmentation network is able to segment defects with consistent accuracy between shallow and deeper defects. On the other hand, Table~\ref{tab:bench_iou} reports the validation and testing IoU values for a U-Net segmentation network trained on latent images from different dimensionality reduction techniques. The U-Net trained on the latent representation of the proposed methodology achieves the highest IoUs, with $0.841$ on validation and $0.836$ on testing. Thereby, segmentation networks trained on the AIRT-Masked-CAAE outperform networks trained on all other baseline representations, including 1D-CNN, with IoUs of $0.763$ and $0.754$, and significantly surpass DAT. Note that the U-Net performs poorly with DAT latent images, as DAT generates unstructured latent spaces where learning-based AIRT approaches struggle to generalize. Compared to C-AET, the U-Net tends to present an IoU increase of 30\% when the latent representation of the AIRT-Masked-CAAE is utilized as input. These results highlight the effectiveness of the AIRT-Masked-CAAE in producing latent representations that not only enhance the clarity of visual defects but also improve downstream learning performance and generalizability. In addition, the proposed masked sequence autoencoding strategy preserves fine-grained defect structures while suppressing background noise, leading to more accurate pixel-wise segmentation. This highlights that the AIRT-Masked-CAAE not only improves defect clarity but also effectively employs its latent representation for downstream automated AIRT.

\subsection{Denoising Evaluation} \label{subsection:denoising}
\begin{figure}[t]
    \centering
    \includegraphics[width=\textwidth]{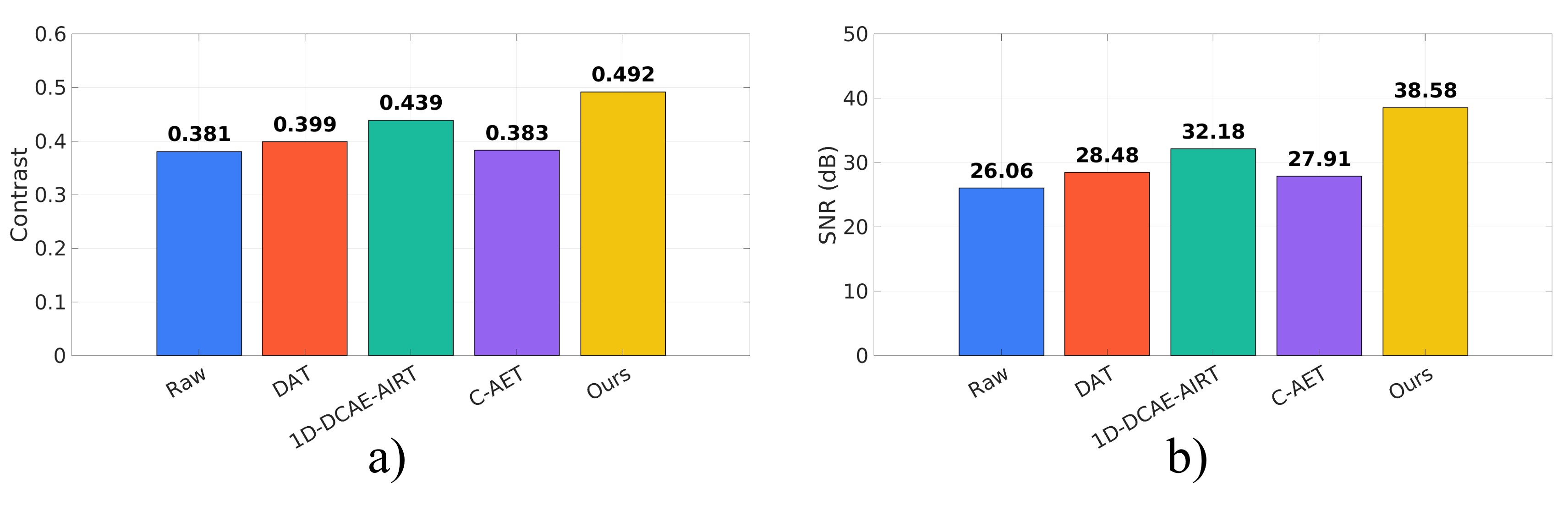}
    \caption{Contrast and SNR of the first 20 PCs, when PCA is applied on the reconstructed, denoised signals using the proposed AIRT-Masked-CAAE, DAT, 1D-DCAE-AIRT, and C-AET.}
    \label{fig:denoising_pca}
    \hfill
\end{figure}

Besides the contrast, SNR, and neural network-based evaluation metrics, the denoising performance of the proposed AIRT-Masked-CAAE is compared against the denoising performance of AIRT AEs; namely, DAT, C-AET, and 1D-DCAE-AIRT. To provide a basis for comparison, the reconstructed signal from each AE is extracted, and then PCA is performed on each reconstructed signal. Consequently, we compute the contrast and SNR for the first 20 PCs. Table \ref{table:pc_comparison} provides a qualitative comparison of the obtained principal components (PCs) after reconstructing the input signal using the aforementioned AEs. Fig. \ref{fig:denoising_pca} compares the signal enhancement in terms of the contrast and SNR provided by PCA when performed on the reconstructed signals from the AEs. Note that an improved contrast and SNR are correlated with improved denoising performance since PCA tends to emphasize dominant variance directions corresponding to meaningful physical or structural features; thus, when the input signal is cleaner, the resulting principal components become more representative of actual defect patterns rather than noise artifacts, leading to clearer and more interpretable PCA visualizations. 

\begin{table*}[t]
\centering
\renewcommand{\arraystretch}{2}
\caption{Qualitative comparisons of different principal components (PCs) after denoising the input thermographic sequence using DAT, 1D-DCAE-AIRT, C-AET, and the proposed AIRT-Masked-CAAE.}
\resizebox{\textwidth}{!}{%
\begin{tabular}{|>{\centering\arraybackslash}m{2.6cm}|>{\centering\arraybackslash}m{4.2cm}|>{\centering\arraybackslash}m{4.2cm}|>{\centering\arraybackslash}m{4.2cm}|}
\hline
\textbf{Method} & \textbf{2nd PC} & \textbf{10th PC} & \textbf{20th PC} \\ \hline

\textbf{Raw} &
{\vspace{3pt}\includegraphics[width=3.25cm, height=2.5cm]{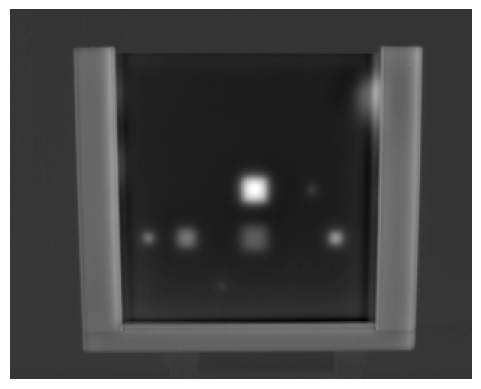}} &
{\vspace{3pt}\includegraphics[width=3.25cm, height=2.5cm]{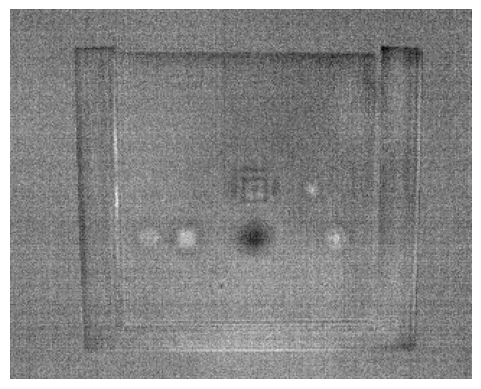}} &
{\vspace{3pt}\includegraphics[width=3.25cm, height=2.5cm]{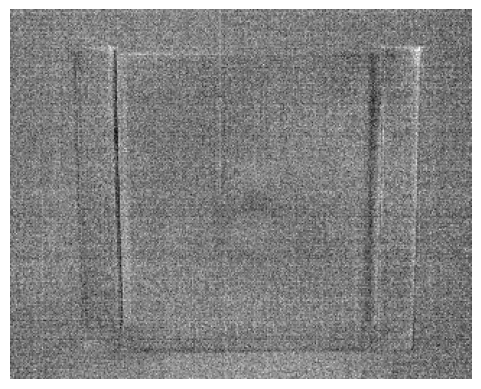}} \\ \hline

\textbf{DAT} &
{\vspace{3pt}\includegraphics[width=3.25cm, height=2.5cm]{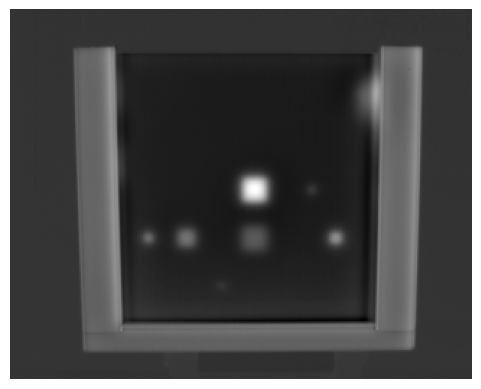}} &
{\vspace{3pt}\includegraphics[width=3.25cm, height=2.5cm]{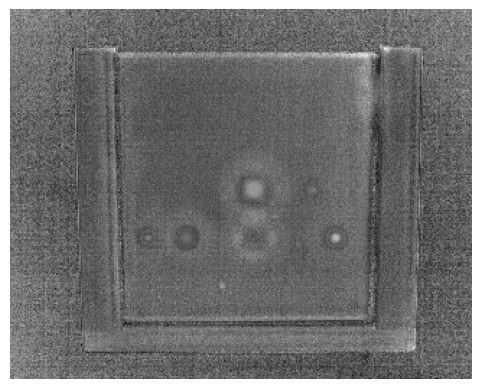}} &
{\vspace{3pt}\includegraphics[width=3.25cm, height=2.5cm]{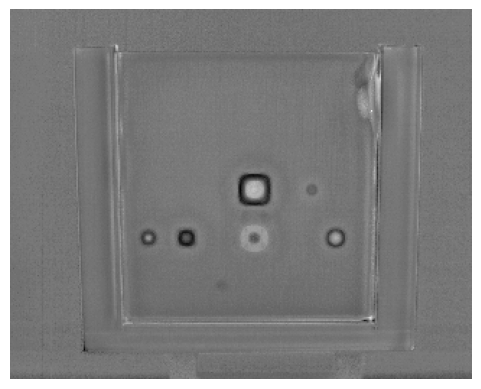}} \\ \hline

\textbf{1D-DCAE-AIRT} &
{\vspace{3pt}\includegraphics[width=3.25cm, height=2.5cm]{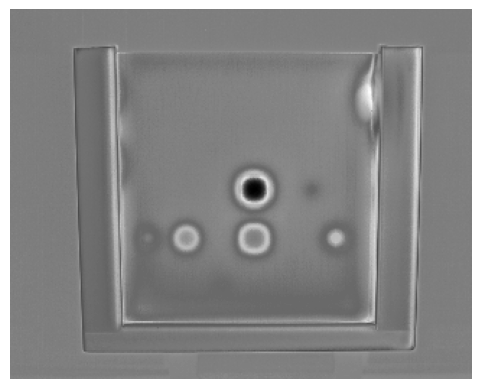}} &
{\vspace{3pt}\includegraphics[width=3.25cm, height=2.5cm]{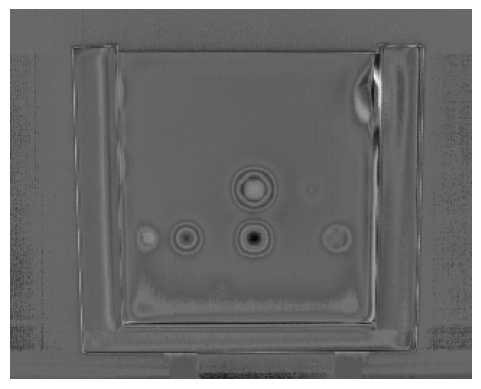}} &
{\vspace{3pt}\includegraphics[width=3.25cm, height=2.5cm]{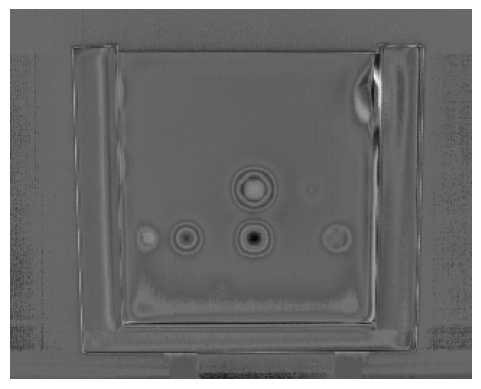}} \\ \hline

\textbf{C-AET} &
{\vspace{3pt}\includegraphics[width=3.25cm, height=2.5cm]{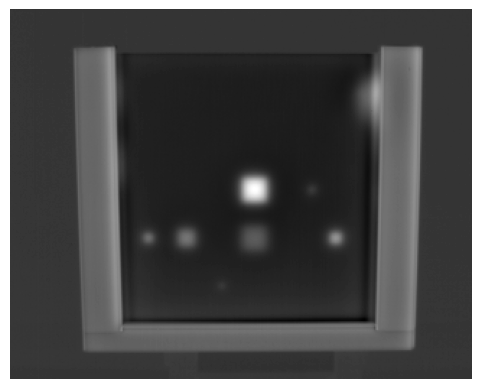}} &
{\vspace{3pt}\includegraphics[width=3.25cm, height=2.5cm]{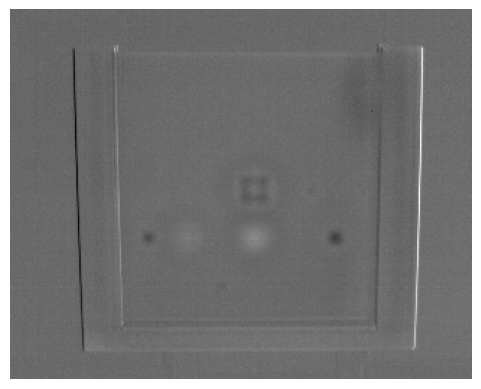}} &
{\vspace{3pt}\includegraphics[width=3.25cm, height=2.5cm]{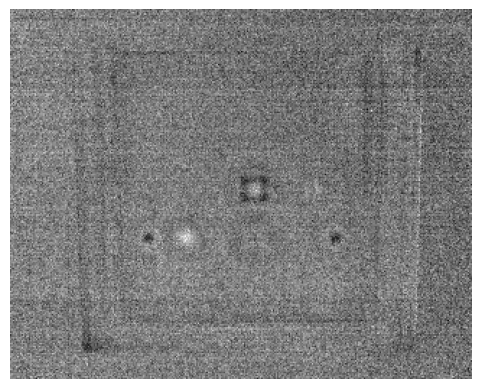}} \\ \hline

\textbf{Ours} &
{\vspace{3pt}\includegraphics[width=3.25cm, height=2.5cm]{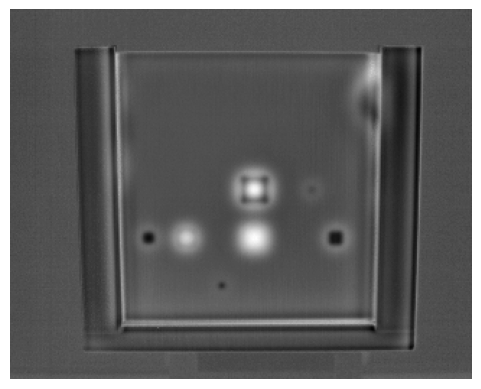}} &
{\vspace{3pt}\includegraphics[width=3.25cm, height=2.5cm]{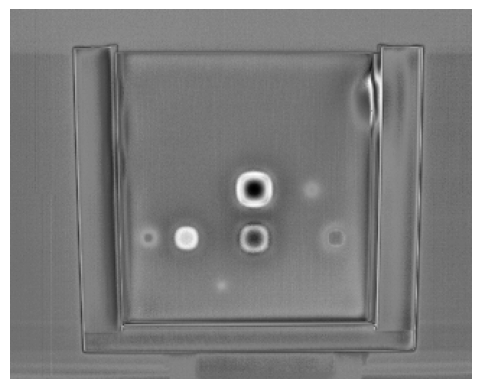}} &
{\vspace{3pt}\includegraphics[width=3.25cm, height=2.5cm]{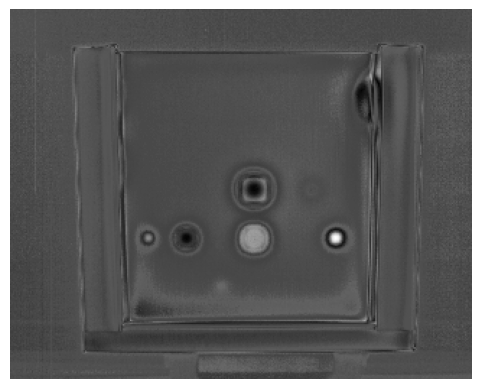}} \\ \hline

\end{tabular}
}
\label{table:pc_comparison}
\end{table*}

\begin{figure}[t]
    \centering
    \includegraphics[width=\textwidth]{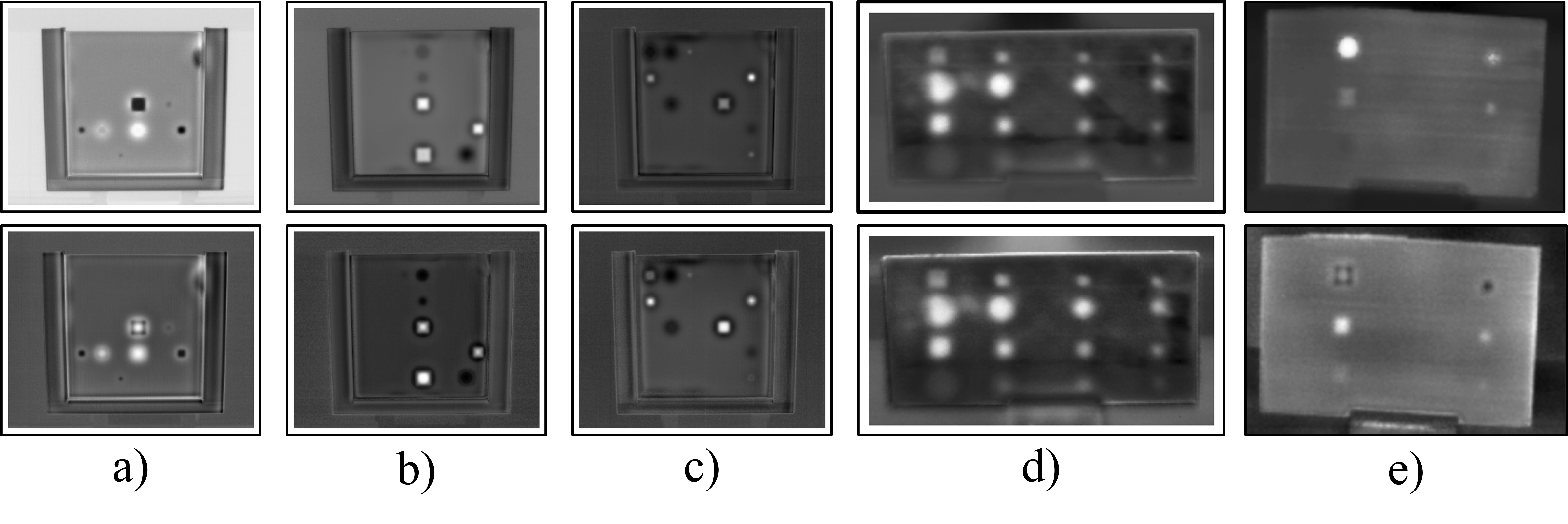}
    \caption{Latent images obtained from AIRT-Masked-CAAE trained on all samples, shown in top row, and trained using the masked sequence autoencoding strategy, shown in bottom row. a), b), and c) are PVC specimens, d) and e) are CFRP and PLA, respectively.}
    \label{fig:ablation_mask}
    \hfill
\end{figure}

\begin{table}[t]
\centering
\caption{Ablation study evaluating the effect of the masked sequence autoencoding strategy on signal enhancement, contrast, and SNR, and neural network-based evaluation, validation, and testing IoU.}
\label{tab:ablation}
\resizebox{\textwidth}{!}{%
\begin{tabular}{|c|c|c|c|c|}
\hline
\textbf{Metric}                                                         & \textbf{Contrast} & \textbf{SNR (dB)} & \textbf{Validation IoU} & \textbf{Testing IoU} \\ \hline
\begin{tabular}[c]{@{}c@{}}AIRT-Masked-CAAE\\ w/o Masking\end{tabular}  & 0.742             & 42.79             & 0.791                   & 0.803                \\ \hline
\begin{tabular}[c]{@{}c@{}}AIRT-Masked-CAAE\\ with Masking\end{tabular} & \textbf{0.791}    & \textbf{48.59}    & \textbf{0.841}          & \textbf{0.836}       \\ \hline
\end{tabular}%
}
\end{table}

\subsection{Ablation Studies} \label{subsection:ablation}
In addition to evaluating the proposed framework for signal enhancement, we conduct a study on the effect of the masked sequence autoencoding strategy, as well as neural network-based and denoising evaluations. Particularly, the efficacy of the masked autoencoding strategy compared to training the network on all pixel responses in terms of signal enhancement metrics, neural network-based evaluation, and training time. Figure \ref{fig:ablation_mask} compares the latent spaces of the proposed CNN-Attention autoencoder trained with and without the masked sequence autoencoding strategy, while Table \ref{tab:ablation} quantifies the contrast, SNR, and IoU for both training schemes on the IRT-PVC dataset. The quantified results in Table \ref{tab:ablation} show that the performance of the AIRT-Masked-CAAE tends to improve when employing the masked sequence training strategy. This is because the masking operation prevents the network from converging to trivial identity reconstruction and instead compels it to learn robust, defect-relevant features that generalize better across samples. Additionally, the model becomes prone to overfitting when trained on all pixel responses. In terms of training time, masking reduces the computational burden since the network does not require processing all pixel responses during each training step, leading to faster convergence and shorter training times. For instance, training the AIRT-Masked-CAAE on all pixel responses takes 18.4 minutes on a laptop grade RTX 3060 GPU. In contrast, when using masked sequence autoencoding with only 1000 samples, the training time is reduced to 36.7 seconds, a factor of $30\times$ improvement, while achieving enhanced dimensionality reduction performance.

\section{Conclusions}
\label{sec: Conclusions}
Active infrared thermography (AIRT) has established itself as an essential technique in NDT for identifying hidden subsurface defects across a range of industrial materials. A challenge that remains in AIRT is the high dimensionality of thermographic sequences and extracting defect-relevant features while suppressing background noise. Classical dimensionality reduction methods, such as PCA and TSR, provide efficient compression but fail to capture non-linear temporal dependencies. Hence, this paper presented the AIRT-Masked-CAAE, a CNN-attention autoencoder enhanced with a masked sequence autoencoding strategy. The proposed framework integrates convolutional feature extraction with attention mechanisms to balance local defect features and long-range dependencies. In addition, the proposed masked sequence autoencoding approach allows the network to focus on defect relevant features, while accelerating training. Extensive experiments on CFRP, PLA, and PVC datasets demonstrated that the AIRT-Masked-CAAE consistently improves contrast and SNR compared to both traditional and state-of-the-art learning-based AIRT dimensionality reduction techniques. Additionally, neural network-based evaluations demonstrated that segmentation models trained on AIRT-Masked-CAAE latent representations outperform their counterparts trained on alternative thermographic representations. Future work will extend the framework to multi-modal thermographic inspections, such as eddy current thermography and vibrothermography. Additionally, future research efforts will be directed towards physics-informed neural network learning to further enhance the physical interpretability of the AIRT-Masked-CAAE.

\section*{Disclosure statement}
About the scientific content of the work, no potential conflict of interest was reported by the author(s).

\section*{Funding}
This work was supported by Khalifa University of Science and Technology under Award 8474000660. The work was also supported by the Advanced Research and Innovation Center (ARIC), which is jointly funded by Aerospace Holding Company LLC, a wholly-owned subsidiary of Mubadala Investment Company PJSC, and Khalifa University for Science and Technology.

\section*{Notes on Contributors}

\noindent \textit{Mohammed Salah} received his BSc. in Mechanical Engineering from the American University of Sharjah, UAE, in 2020 and his MSc. in Mechanical Engineering from Khalifa University, Abu Dhabi, UAE, in 2022. He is currently pursuing his Ph.D. in Robotics at Khalifa University, Abu Dhabi, UAE. His research interests include robotic automation, thermography, and non-destructive testing.

\noindent \textit{Eman Ouda} is a Postdoctoral Fellow in the Aerospace Engineering Department at Khalifa University, Abu Dhabi, UAE. She received her Ph.D. in Management Science and Engineering from Khalifa University in 2025 and her BSc. in Electrical Engineering from the American University of Sharjah, UAE, in 2019. Her research interests include stochastic simulation, resilience engineering, and simulation-based optimization in healthcare operations, as well as robotic automation, thermography, and non-destructive testing, bridging applications across healthcare and aerospace domains.

\noindent \textit{Stefano Sfarra} Received the Ph.D. degree in mechanical, management, and energy engineering from the University of L’Aquila (UNIVAQ), L’Aquila, Italy, in 2011. He was a Postdoctoral Fellow with the University of L’Aquila until October 2017, when he became a Researcher with a fixed-term contract with the Department of Industrial and Information Engineering and Economics, UNIVAQ. From 2020, he is an Associate Professor. Prof. Sfarra has been included from 2020 to 2023 in the list of "World's Top 2\% of Scientists" edited by Stanford University. He has authored or coauthored more than 270 papers in journals and international conferences. His current research interests include infrared thermography, heat transfer, bio-thermodynamics, composites, energy saving, and finite element modelling.

\noindent \textit{Davor Svetinovic} is a professor at the Department of Computer Science, Khalifa University, Abu Dhabi, UAE. He received his doctorate in computer science from the University of Waterloo, Waterloo, ON, Canada, in 2006. Previously, he worked at WU Wien, TU Wien, Austria, and Lero -- the Irish Software Engineering Center, Ireland. He was a visiting professor and a research affiliate at MIT and MIT Media Lab, MIT, USA. Davor has extensive experience working on complex multidisciplinary research projects. He has published over 100 papers in leading journals and conferences and is a highly cited researcher in blockchain technology. His research interests include cybersecurity, blockchain technology, crypto-economics, trust, and software engineering. His career has furthered his interest and expertise in developing advanced research capabilities and institutions in emerging economies. He is a Senior Member of IEEE and ACM (Lifetime), and an affiliate of the Mohammed Bin Rashid Academy of Scientists.

\noindent \textit{Yusra Abdulrahman} received the B.Sc. degree from The University of
Arizona, in 2014, and the M.Sc. and Ph.D. degrees
from the Massachusetts Institute of Technology and
Masdar Institute of Science and Technology
Cooperative Program (MIT and MICP), in
2016 and 2020, respectively. She is currently the Deputy Director of the Advanced Research and Innovation Center, KU, and an Assistant Professor at the Department of
Aerospace Engineering, Khalifa University. Her
expertise lies in robotics, artificial intelligence
(AI), and non-destructive testing (NDT). She has received awards from the
UAE Ministry of Energy and Industry for her research contributions.

\bibliographystyle{tfq}
\bibliography{main.bib}

@article{venegas2021ndt,
  title={NDT inspection of aeronautical components by projected thermal diffusivity analysis},
  author={Venegas, P and Peran, J and Usamentiaga, R and S{\'a}ez de Oc{\'a}riz, I},
  journal={Quantitative InfraRed Thermography Journal},
  volume={18},
  number={1},
  pages={34--49},
  year={2021},
  publisher={Taylor \& Francis}
}

@article{helvig2025automated,
  title={Automated crack detection on metallic materials with flying-spot thermography using deep learning and progressive training},
  author={Helvig, Kevin and Trouv{\'e}-Peloux, Pauline and Gaverina, Ludovic and Abeloos, Baptiste and Roche, Jean-Michel},
  journal={Quantitative InfraRed Thermography Journal},
  volume={22},
  number={1},
  pages={21--40},
  year={2025},
  publisher={Taylor \& Francis}
}

@article{oswald2025inductive,
  title={Inductive thermography--review of a non-destructive inspection technique for surface crack detection},
  author={Oswald-Tranta, Beate},
  journal={Quantitative InfraRed Thermography Journal},
  pages={1--25},
  year={2025},
  publisher={Taylor \& Francis}
}

@article{alhammad2024multi,
  title={Multi-label classification algorithms for composite materials under infrared thermography testing},
  author={Alhammad, Muflih and Avdelidis, Nicolas Peter and Ibarra Castanedo, Clemente and Maldague, Xavier and Zolotas, Argyrios and Torbali, Ebubekir and Genest, Marc},
  journal={Quantitative InfraRed Thermography Journal},
  volume={21},
  number={1},
  pages={3--29},
  year={2024},
  publisher={Taylor \& Francis}
}

@article{hsiao2025two,
  title={Two-dimensional Hilbert-Huang transform-based thermographic data processing for non-destructive material defect detection},
  author={Hsiao, Tung-Yu and Sfarra, Stefano and Liu, Yi and Yao, Yuan},
  journal={Quantitative InfraRed Thermography Journal},
  volume={22},
  number={4},
  pages={297--312},
  year={2025},
  publisher={Taylor \& Francis}
}

@article{mayr2017non,
  title={Non-destructive testing procedure for porosity determination in carbon fibre reinforced plastics using pulsed thermography},
  author={Mayr, G and Gresslehner, KH and Hendorfer, G},
  journal={Quantitative InfraRed Thermography Journal},
  volume={14},
  number={2},
  pages={263--274},
  year={2017},
  publisher={Taylor \& Francis}
}

@article{bardhan2022designing,
  title={Designing of an inflammatory knee joint thermogram dataset for arthritis classification using deep convolution neural network.},
  author={Bardhan, Shawli and Nath, Satyabrata and Debnath, Tathagata and Bhattacharjee, Debotosh and Bhowmik, Mrinal Kanti},
  journal={Quantitative InfraRed Thermography Journal},
  volume={19},
  number={3},
  pages={145--171},
  year={2022},
  publisher={Taylor \& Francis}
}

@article{mueller2022defect,
  title={Defect shape detection and defect reconstruction in active thermography by means of two-dimensional convolutional neural network as well as spatiotemporal convolutional LSTM network},
  author={Mueller, David and Netzelmann, Udo and Valeske, Bernd},
  journal={Quantitative InfraRed Thermography Journal},
  volume={19},
  number={2},
  pages={126--144},
  year={2022},
  publisher={Taylor \& Francis}
}

@article{criniere2014inverse,
  title={Inverse model for defect characterisation of externally glued CFRP on reinforced concrete structures: comparative study of square pulsed and pulsed thermography},
  author={Crini{\`e}re, Antoine and Dumoulin, Jean and Ibarra-Castanedo, Clemente and Maldague, Xavier},
  journal={Quantitative InfraRed Thermography Journal},
  volume={11},
  number={1},
  pages={84--114},
  year={2014},
  publisher={Taylor \& Francis}
}

@article{sun2013analysis,
  title={Analysis of data processing methods for pulsed thermal imaging characterisation of delaminations},
  author={Sun, Jiangang},
  journal={Quantitative InfraRed Thermography Journal},
  volume={10},
  number={1},
  pages={9--25},
  year={2013},
  publisher={Taylor \& Francis}
}

@article{zhang2025automatic,
  title={Automatic segmentation of microporous defects in composite film materials based on the improved attention U-Net module},
  author={Zhang, Zhiyang and Hu, Jue and Wang, Rongbang and Chen, Xueyan and Yang, Dazhi and Vavilov, Vladimir P and Duan, Yuxia and Zhang, Hai},
  journal={Quantitative InfraRed Thermography Journal},
  volume={22},
  number={4},
  pages={313--328},
  year={2025},
  publisher={Taylor \& Francis}
}

@Article{Tai2025,
AUTHOR = {Tai, Jan Lean and Sultan, Mohamed Thariq Hameed and Łukaszewicz, Andrzej and Józwik, Jerzy and Oksiuta, Zbigniew and Shahar, Farah Syazwani},
TITLE = {Recent Trends in Non-Destructive Testing Approaches for Composite Materials: A Review of Successful Implementations},
JOURNAL = {Materials},
VOLUME = {18},
YEAR = {2025},
NUMBER = {13},
ARTICLE-NUMBER = {3146},
URL = {https://www.mdpi.com/1996-1944/18/13/3146},
PubMedID = {40649634},
ISSN = {1996-1944},
DOI = {10.3390/ma18133146}
}

@article{Carosena2004,
doi = {10.1088/0957-0233/15/9/R01},
url = {https://dx.doi.org/10.1088/0957-0233/15/9/R01},
year = {2004},
month = {jul},
publisher = {},
volume = {15},
number = {9},
pages = {R27},
author = {Carosena Meola and Giovanni M Carlomagno},
title = {Recent advances in the use of infrared thermography},
journal = {Measurement Science and Technology},
}

@Article{ma2025,
AUTHOR = {Tai, Jan Lean and Sultan, Mohamed Thariq Hameed and Łukaszewicz, Andrzej and Józwik, Jerzy and Oksiuta, Zbigniew and Shahar, Farah Syazwani},
TITLE = {Recent Trends in Non-Destructive Testing Approaches for Composite Materials: A Review of Successful Implementations},
JOURNAL = {Materials},
VOLUME = {18},
YEAR = {2025},
NUMBER = {13},
ARTICLE-NUMBER = {3146},
URL = {https://www.mdpi.com/1996-1944/18/13/3146},
PubMedID = {40649634},
ISSN = {1996-1944},
DOI = {10.3390/ma18133146}
}

@article{TOWS2020,
title = {Successes and challenges in non-destructive testing of aircraft composite structures},
journal = {Chinese Journal of Aeronautics},
volume = {33},
number = {3},
pages = {771-791},
year = {2020},
issn = {1000-9361},
doi = {https://doi.org/10.1016/j.cja.2019.09.017},
url = {https://www.sciencedirect.com/science/article/pii/S1000936119303474},
author = {Hossein TOWSYFYAN and Ander BIGURI and Richard BOARDMAN and Thomas BLUMENSATH}
}

@article{DUA2021,
title = {InfraRed image correlation for non-destructive testing and evaluation of delaminations in glass fibre reinforced polymer materials},
journal = {Infrared Physics \& Technology},
volume = {116},
pages = {103803},
year = {2021},
issn = {1350-4495},
doi = {https://doi.org/10.1016/j.infrared.2021.103803},
url = {https://www.sciencedirect.com/science/article/pii/S1350449521001754},
author = {Geetika Dua and Ravibabu Mulaveesala and Priyanka Mishra and Jasleen kaur},
}

@article{AIRT2015,
author = {Lizaranzu Fernández, Miguel and Lario, Alberto and Chiminelli, A. and Amenabar, Iban},
year = {2015},
month = {03},
pages = {},
title = {Non-Destructive Testing of Composite Materials by means of Active Thermography-Based Tools},
volume = {71},
journal = {Infrared Physics \& Technology},
doi = {10.1016/j.infrared.2015.02.006}
}

@article{irt_survey,
title = {Infrared machine vision and infrared thermography with deep learning: A review},
journal = {Infrared Physics \& Technology},
volume = {116},
pages = {103754},
year = {2021},
issn = {1350-4495},
doi = {https://doi.org/10.1016/j.infrared.2021.103754},
url = {https://www.sciencedirect.com/science/article/pii/S1350449521001262},
author = {Yunze He and Baoyuan Deng and Hongjin Wang and Liang Cheng and Ke Zhou and Siyuan Cai and Francesco Ciampa},
}

@inproceedings{thermosense_attention,
author = {Bardia Yousefi and Xavier P.V. Maldague},
title = {{Attention mechanism for computational thermography}},
volume = {13047},
booktitle = {Thermosense: Thermal Infrared Applications XLVI},
editor = {Fernando L{\'o}pez and Nicolas P. Avdelidis and Giovanni Ferrarini},
organization = {International Society for Optics and Photonics},
publisher = {SPIE},
pages = {130471A},
year = {2024},
doi = {10.1117/12.3013999},
URL = {https://doi.org/10.1117/12.3013999}
}

@INPROCEEDINGS{tsr_2,
  author={Bo, Chunqiang and Hu, Hong and Lei, Guobin and Liu, Ze and Shao, Junhao},
  booktitle={2018 IEEE International Conference on Information and Automation (ICIA)}, 
  title={Non-destructive Testing of Airfoil Based on Infrared Lock-in Thermography}, 
  year={2018},
  volume={},
  number={},
  pages={1623-1628},
  doi={10.1109/ICInfA.2018.8812574}}

@ARTICLE{pca_2,
  author={Wen, Ching-Mei and Sfarra, Stefano and Gargiulo, Gianfranco and Yao, Yuan},
  journal={IEEE Transactions on Industrial Informatics}, 
  title={Thermographic Data Analysis for Defect Detection by Imposing Spatial Connectivity and Sparsity Constraints in Principal Component Thermography}, 
  year={2021},
  volume={17},
  number={6},
  pages={3901-3909},
  doi={10.1109/TII.2020.3010273}}

@article{pca_3,
title = {Low-rank sparse principal component thermography (sparse-PCT): Comparative assessment on detection of subsurface defects},
journal = {Infrared Physics \& Technology},
volume = {98},
pages = {278-284},
year = {2019},
issn = {1350-4495},
doi = {https://doi.org/10.1016/j.infrared.2019.03.012},
url = {https://www.sciencedirect.com/science/article/pii/S1350449518308727},
author = {Bardia Yousefi and Stefano Sfarra and Fabrizio Sarasini and Clemente Ibarra Castanedo and Xavier P.V. Maldague},
}

@article{ppt_1,
title = {Modified pulse-phase thermography algorithms for improved contrast-to-noise ratio from pulse-excited thermographic sequences},
journal = {NDT \& E International},
volume = {116},
pages = {102325},
year = {2020},
issn = {0963-8695},
doi = {https://doi.org/10.1016/j.ndteint.2020.102325},
url = {https://www.sciencedirect.com/science/article/pii/S0963869519307546},
author = {Udo Netzelmann and David Müller},
}

@inproceedings{thermosense_concrete,
author = {Sandra Pozzer and Gabriel Ramos and Ehsan Rezazadeh Azar and Ahmad Osman and Ahmed El Refai and Fernando L{\'o}pez and Clemente Ibarra-Castanedo and Xavier Maldague},
title = {{A few-shot learning approach for the segmentation of subsurface defects in thermography images of concrete structures}},
volume = {13047},
booktitle = {Thermosense: Thermal Infrared Applications XLVI},
editor = {Fernando L{\'o}pez and Nicolas P. Avdelidis and Giovanni Ferrarini},
organization = {International Society for Optics and Photonics},
publisher = {SPIE},
pages = {130470F},
year = {2024},
doi = {10.1117/12.3013684},
URL = {https://doi.org/10.1117/12.3013684}
}

@Article{unet_study,
AUTHOR = {Fang, Qiang and Ibarra-Castanedo, Clemente and Garrido, Iván and Duan, Yuxia and Maldague, Xavier},
TITLE = {Automatic Detection and Identification of Defects by Deep Learning Algorithms from Pulsed Thermography Data},
JOURNAL = {Sensors},
VOLUME = {23},
YEAR = {2023},
NUMBER = {9},
ARTICLE-NUMBER = {4444},
URL = {https://www.mdpi.com/1424-8220/23/9/4444},
PubMedID = {37177648},
ISSN = {1424-8220},
DOI = {10.3390/s23094444}
}

@inproceedings{thermosense_artwork,
author = {Masashi Ishikawa and Stefano Sfarra and Panagiotis Theodorakeas},
title = {{Active thermography non-destructive inspection of a damaged artwork with a complex shape}},
volume = {13047},
booktitle = {Thermosense: Thermal Infrared Applications XLVI},
editor = {Fernando L{\'o}pez and Nicolas P. Avdelidis and Giovanni Ferrarini},
organization = {International Society for Optics and Photonics},
publisher = {SPIE},
pages = {1304717},
year = {2024},
doi = {10.1117/12.3013386},
URL = {https://doi.org/10.1117/12.3013386}
}

@Article{csk_nvs,
author={Salah, Mohammed
and Ayyad, Abdulla
and Ramadan, Mohammed
and Abdulrahman, Yusra
and Swart, Dewald
and Abusafieh, Abdelqader
and Seneviratne, Lakmal
and Zweiri, Yahya},
title={High speed neuromorphic vision-based inspection of countersinks in automated manufacturing processes},
journal={Journal of Intelligent Manufacturing},
year={2023},
month={Aug},
day={18},
issn={1572-8145},
doi={10.1007/s10845-023-02187-0},
url={https://doi.org/10.1007/s10845-023-02187-0}
}

@article{attention_unet,
author = {Guangyu Zhou and Zhijie Zhang and Wuliang Yin and Haoze Chen and Luxiang Wang and Dong Wang and Huidong Ma},
title = {Surface defect detection of CFRP materials based on infrared thermography and Attention U-Net algorithm},
journal = {Nondestructive Testing and Evaluation},
volume = {39},
number = {2},
pages = {238--257},
year = {2024},
publisher = {Taylor \& Francis},
doi = {10.1080/10589759.2023.2191954},
URL = { 
        https://doi.org/10.1080/10589759.2023.2191954
},
eprint = { 
        https://doi.org/10.1080/10589759.2023.2191954
}
}

@article{flexible_framework,
title = {A flexible deep learning framework for thermographic inspection of composites},
journal = {NDT \& E International},
volume = {139},
pages = {102926},
year = {2023},
issn = {0963-8695},
doi = {https://doi.org/10.1016/j.ndteint.2023.102926},
url = {https://www.sciencedirect.com/science/article/pii/S096386952300141X},
author = {Zongfei Tong and Liangliang Cheng and Shejuan Xie and Mathias Kersemans},
}

@ARTICLE{3d_cnn,
  author={Dong, Yafei and Xia, Chenjie and Yang, Jiangxin and Cao, Yanlong and Cao, Yanpeng and Li, Xin},
  journal={IEEE Transactions on Industrial Informatics}, 
  title={Spatio-Temporal 3-D Residual Networks for Simultaneous Detection and Depth Estimation of CFRP Subsurface Defects in Lock-In Thermography}, 
  year={2022},
  volume={18},
  number={4},
  pages={2571-2581},
  doi={10.1109/TII.2021.3103019}}

@Article{tsr,
AUTHOR = {Schager, Alexander and Zauner, Gerald and Mayr, Günther and Burgholzer, Peter},
TITLE = {Extension of the Thermographic Signal Reconstruction Technique for an Automated Segmentation and Depth Estimation of Subsurface Defects},
JOURNAL = {Journal of Imaging},
VOLUME = {6},
YEAR = {2020},
NUMBER = {9},
ARTICLE-NUMBER = {96},
URL = {https://www.mdpi.com/2313-433X/6/9/96},
PubMedID = {34460753},
ISSN = {2313-433X},
DOI = {10.3390/jimaging6090096}
}

@Article{pt_dataset,
AUTHOR = {Wei, Ziang and Osman, Ahmad and Valeske, Bernd and Maldague, Xavier},
TITLE = {A Dataset of Pulsed Thermography for Automated Defect Depth Estimation},
JOURNAL = {Applied Sciences},
VOLUME = {13},
YEAR = {2023},
NUMBER = {24},
ARTICLE-NUMBER = {13093},
URL = {https://www.mdpi.com/2076-3417/13/24/13093},
ISSN = {2076-3417},
DOI = {10.3390/app132413093}
}

@article{cnnlstm,
author = {David Müller, Udo Netzelmann and Bernd Valeske},
title = {Defect shape detection and defect reconstruction in active thermography by means of two-dimensional convolutional neural network as well as spatiotemporal convolutional LSTM network},
journal = {Quantitative InfraRed Thermography Journal},
volume = {19},
number = {2},
pages = {126--144},
year = {2022},
publisher = {Taylor \& Francis},
doi = {10.1080/17686733.2020.1810883},


URL = { 
    
        https://doi.org/10.1080/17686733.2020.1810883
    
    

},
eprint = { 
    
        https://doi.org/10.1080/17686733.2020.1810883
    
    

}

}

@Article{ndt_review,
AUTHOR = {Torbali, Muhammet E. and Zolotas, Argyrios and Avdelidis, Nicolas P.},
TITLE = {A State-of-the-Art Review of Non-Destructive Testing Image Fusion and Critical Insights on the Inspection of Aerospace Composites towards Sustainable Maintenance Repair Operations},
JOURNAL = {Applied Sciences},
VOLUME = {13},
YEAR = {2023},
NUMBER = {4},
ARTICLE-NUMBER = {2732},
URL = {https://www.mdpi.com/2076-3417/13/4/2732},
ISSN = {2076-3417},
DOI = {10.3390/app13042732}
}

@Article{yusra_taguchi,
author={Abdulrahman, Y. A.
and Omar, M. A.
and Said, Z.
and Obeideli, F.
and Abusafieh, A.
and Sankaran, G. N.},
title={A Taguchi Design of Experiment Approach to Pulse and Lock in Thermography, Applied to CFRP Composites},
journal={Journal of Nondestructive Evaluation},
year={2017},
month={Oct},
day={04},
volume={36},
number={4},
pages={72},
issn={1573-4862},
doi={10.1007/s10921-017-0450-4},
url={https://doi.org/10.1007/s10921-017-0450-4}
}

@article{construction_ndt,
title = {Deep learning augmented infrared thermography for unmanned aerial vehicles structural health monitoring of roadways},
journal = {Automation in Construction},
volume = {148},
pages = {104784},
year = {2023},
issn = {0926-5805},
doi = {https://doi.org/10.1016/j.autcon.2023.104784},
url = {https://www.sciencedirect.com/science/article/pii/S0926580523000444},
author = {Nitin Nagesh Kulkarni and Koosha Raisi and Nicholas A. Valente and Jason Benoit and Tzuyang Yu and Alessandro Sabato},
}

@article{construction_uav,
author = {Wang, Jiehui and Ueda, Tamon},
year = {2023},
month = {01},
pages = {},
title = {A review study on unmanned aerial vehicle and mobile robot technologies on damage inspection of reinforced concrete structures},
volume = {24},
journal = {Structural Concrete},
doi = {10.1002/suco.202200846}
}

@Article{pct,
AUTHOR = {Ebrahimi, Samira and Fleuret, Julien and Klein, Matthieu and Théroux, Louis-Daniel and Georges, Marc and Ibarra-Castanedo, Clemente and Maldague, Xavier},
TITLE = {Robust Principal Component Thermography for Defect Detection in Composites},
JOURNAL = {Sensors},
VOLUME = {21},
YEAR = {2021},
NUMBER = {8},
ARTICLE-NUMBER = {2682},
URL = {https://www.mdpi.com/1424-8220/21/8/2682},
PubMedID = {33920261},
ISSN = {1424-8220},
DOI = {10.3390/s21082682}
}

@ARTICLE{pct_gaussian,
  author={Liu, Wei and Hou, Beiping and Yao, Yuan and Zhou, Le},
  journal={IEEE Access}, 
  title={Signal Enhancement in Defect Detection of CFRP Material Using a Combination of Difference of Gaussian Convolutions and Sparse Principal Component Thermography}, 
  year={2022},
  volume={10},
  number={},
  pages={108103-108116},
  doi={10.1109/ACCESS.2022.3212538}}

@article{sparse_kernel_pca,
title = {Analyzing a series of thermal infrared images to identify defects using a hybrid approach that combines robust principal component analysis and image segmentation},
journal = {NDT \& E International},
volume = {137},
pages = {102818},
year = {2023},
issn = {0963-8695},
doi = {https://doi.org/10.1016/j.ndteint.2023.102818},
url = {https://www.sciencedirect.com/science/article/pii/S0963869523000336},
author = {Yishuo Huang and Chin-Lung Chen and Chih-Hung Chiang},
}

@Article{irt_depth,
AUTHOR = {Wei, Ziang and Osman, Ahmad and Valeske, Bernd and Maldague, Xavier},
TITLE = {A Dataset of Pulsed Thermography for Automated Defect Depth Estimation},
JOURNAL = {Applied Sciences},
VOLUME = {13},
YEAR = {2023},
NUMBER = {24},
ARTICLE-NUMBER = {13093},
URL = {https://www.mdpi.com/2076-3417/13/24/13093},
ISSN = {2076-3417},
DOI = {10.3390/app132413093}
}

@article{autoencoder,
author = {Yi Liu and Fumin Wang and Kaixin Liu and Miranda Mostacci and Yuan Yao and Stefano Sfarra},
title = {Deep convolutional autoencoder thermography for artwork defect detection},
journal = {Quantitative InfraRed Thermography Journal},
volume = {0},
number = {0},
pages = {1--17},
year = {2023},
publisher = {Taylor \& Francis},
doi = {10.1080/17686733.2023.2225246},


URL = { 
    
        https://doi.org/10.1080/17686733.2023.2225246
    
    

},
eprint = { 
    
        https://doi.org/10.1080/17686733.2023.2225246
    
    

}

}

@article{1d_cnn,
author = {Zhang, Yubin and Xu, Changhang and Liu, Pengqian and Xie, Jing and Han, Yage and Liu, Rui and Chen, Lina},
year = {2024},
month = {03},
pages = {111216},
title = {One-dimensional deep convolutional autoencoder active infrared thermography: Enhanced visualization of internal defects in FRP composites},
journal = {Composites Part B Engineering},
doi = {10.1016/j.compositesb.2024.111216}
}

@article{yusra_3,
author = {Omar, Mohammed and Said, Zafar and Raisi, A and Abdulrahman, Yusra and Abusafieh, A and Sankaran, GN},
year = {2016},
month = {08},
pages = {51},
title = {The Calibration and Sensitivity Aspects of a Self-Referencing Routine When Applied to Composites Inspection: Using a Pulsed Thermographic Setup},
volume = {35},
journal = {Journal of Nondestructive Evaluation},
doi = {10.1007/s10921-016-0367-3}
}

@Article{cfrp_deep,
AUTHOR = {Wei, Ziang and Fernandes, Henrique and Herrmann, Hans-Georg and Tarpani, Jose Ricardo and Osman, Ahmad},
TITLE = {A Deep Learning Method for the Impact Damage Segmentation of Curve-Shaped CFRP Specimens Inspected by Infrared Thermography},
JOURNAL = {Sensors},
VOLUME = {21},
YEAR = {2021},
NUMBER = {2},
ARTICLE-NUMBER = {395},
URL = {https://www.mdpi.com/1424-8220/21/2/395},
PubMedID = {33429939},
ISSN = {1424-8220},
DOI = {10.3390/s21020395}
}

@misc{pca_guided_ae,
      title={PCA-Guided Autoencoding for Structured Dimensionality Reduction in Active Infrared Thermography}, 
      author={Mohammed Salah and Numan Saeed and Davor Svetinovic and Stefano Sfarra and Mohammed Omar and Yusra Abdulrahman},
      year={2025},
      eprint={2508.07773},
      archivePrefix={arXiv},
      primaryClass={eess.IV},
      url={https://arxiv.org/abs/2508.07773}, 
}

@ARTICLE{dat,
  author={Liu, Kaixin and Zheng, Mingkai and Liu, Yi and Yang, Jianguo and Yao, Yuan},
  journal={IEEE Transactions on Industrial Informatics}, 
  title={Deep Autoencoder Thermography for Defect Detection of Carbon Fiber Composites}, 
  year={2023},
  volume={19},
  number={5},
  pages={6429-6438},
  doi={10.1109/TII.2022.3172902}}

@ARTICLE{constrained_ae,
  author={Kaur, Kirandeep and Mulaveesala, Ravibabu and Mishra, Priyanka},
  journal={IEEE Sensors Journal}, 
  title={Constrained Autoencoder-Based Pulse Compressed Thermal Wave Imaging for Sub-Surface Defect Detection}, 
  year={2022},
  volume={22},
  number={18},
  pages={17335-17342},
  doi={10.1109/JSEN.2021.3056394}}

\end{document}